\documentclass[preprint,12pt]{elsarticle}

\usepackage{tikz}
\usepackage{amsmath}
\usepackage{graphicx}
\usepackage{comment}
\usepackage{array,multirow,graphicx,xcolor}
\usepackage{float}
\usepackage{amssymb}
\usepackage{booktabs}
\usepackage{enumitem}
\usepackage{multicol}
\usepackage{makecell}
\usepackage{filecontents}
\usepackage[font=small,skip=0pt]{caption}
\usepackage{subcaption}
\usepackage{ifthen}
\usepackage{gensymb}
\usepackage{wrapfig}
\usepackage{setspace}
\usepackage[export]{adjustbox}

\usepackage[]{xurl}
\usepackage{hyperref}

\usepackage{ifthen}

\newboolean{anony} 
\setboolean{anony}{false} 
\newboolean{12pager} 
\setboolean{12pager}{true}
\newboolean{journal}
\setboolean{journal}{true}

\newcommand\hz[1]{\textcolor{orange}{#1 \ \ - HZ}}

\newcommand{\ara}{ARA} 

\newcommand{\araran}{AraRAN} 
\newcommand{\aratvws}{AraMIMO-tvws} 
\newcommand{\arac}{AraMIMO-c} 
\newcommand{\aramm}{AraMIMO-mm} 
\newcommand{\arasdr}{AraSDR} 

\newcommand{\arahaul}{AraHaul} 
\newcommand{\aviatmicro}{AraHaul-micro} 
\newcommand{\aviatmm}{AraHaul-mm} 
\newcommand{\araoptical}{AraHaul-fso} 
\newcommand{\arantn}{AraHaul-leo} 

\newcommand{\arasoft}{AraSoft} 
\newcommand{\arahost}{AraHost} 
\newcommand{\aracontroller}{AraController} 
\newcommand{\aramanagement}{AraManagement} 

\newcommand{\ISU}{Iowa State University} 
\newcommand{\ames}{City of Ames} 
\newcommand{\boone}{Boone} 
\newcommand{\gilbert}{Gilbert} 
\newcommand{\nevada}{Nevada} 
\newcommand{\McCallsburg}{McCallsburg} 
\newcommand{\statecenter}{State Center} 
\newcommand{\isics}{ISICS Tower} 
\newcommand{\infobunker}{InfoBunker} 

\newcommand{\iowa}{Iowa} 
\newcommand{\agronomyfarm}{Agronomy Farm} 
\newcommand{\curtissfarm}{Curtiss Farm} 
\newcommand{\wilsonhall}{Wilson Hall} 
\newcommand{\researchpark}{Research Park} 

\newcommand{\subHeading}[1]{
\vspace*{0.1cm}
\noindent \textbf{#1}
}

\journal{Elevier Computer Networks}

\begin{document}

\begin{frontmatter}



\title{Design and Implementation of ARA Wireless Living Lab \\ for Rural Broadband and Applications}


\author[iastate]{Taimoor Ul Islam\corref{cor1}}
\ead{tislam@iastate.edu}
\cortext[cor1]{Corresponding Author}
\author[iastate]{Joshua Ofori Boateng}
\author[iastate]{Md Nadim}
\author[iastate]{Guoying Zu}
\author[iastate]{Mukaram Shahid}
\author[uci]{Xun Li}
\author[iastate]{Tianyi Zhang}
\author[osu]{Salil Reddy}
\author[iastate]{Wei Xu}
\author[uci]{Ataberk Atalar}
\author[iastate]{Vincent Lee}
\author[osu]{Yung-Fu Chen}
\author[iastate]{Evan Gossling}
\author[iastate]{Elisabeth Permatasari}
\author[iastate]{Christ Somiah}
\author[iastate]{Owen Perrin}
\author[iastate]{Zhibo Meng}
\author[iastate]{Reshal Afzal}
\author[iastate]{Sarath Babu}
\author[iastate]{Mohammed Soliman}
\author[iastate]{Ali Hussain}
\author[iastate]{Daji Qiao}
\author[iastate]{Mai Zheng}
\author[uci]{Ozdal Boyraz}
\author[iastate]{Yong Guan}
\author[osu]{Anish Arora}
\author[iastate]{Mohamed Y. Selim}
\author[iastate]{Arsalan Ahmad}
\author[iastate]{Myra B. Cohen}
\author[bitripple]{Mike Luby}
\author[microsoft]{Ranveer Chandra}
\author[kth]{James Gross}
\author[anl]{Kate Keahey}
\author[iastate]{Hongwei Zhang}

\affiliation[iastate]{organization={Iowa State University}} 
\affiliation[uci]{organization={University of California, Irvine}}
\affiliation[osu]{organization={Ohio State University}}
\affiliation[kth]{organization={KTH Royal Institute of Technology}}
\affiliation[anl]{organization={Argonne National Laboratory}}
\affiliation[microsoft]{organization={Microsoft Research}}
\affiliation[bitripple]{country={BitRipple Inc.}}


\begin{abstract}
Addressing the broadband gap between rural and urban regions requires rural-focused wireless research and innovation. In the meantime, rural regions provide rich, diverse use cases of advanced wireless, and they offer unique real-world settings for piloting applications that advance the frontiers of wireless systems (e.g., teleoperation of ground and aerial vehicles). 
    To fill the broadband gap 
    and to leverage the unique opportunities that rural regions provide for piloting advanced wireless applications, we design and implement the \ara\ wireless living lab for research and innovation in rural wireless systems and their applications in precision agriculture, community services, and so on. 
\ara\ focuses on the unique community, application, and economic context of rural regions, and it features the first-of-its-kind, real-world deployment of long-distance, high-capacity terrestrial wireless x-haul and access platforms as well as low-earth-orbit (LEO) satellite communications platforms across a rural area of diameter over 30\,km. 
    The high-capacity x-haul platforms operate at the 11\,GHz, 14\,GHz, 71--86\,GHz, and 194\,THz bands and offer communication capacities of up to 160\,Gbps at per-hop distances up to 15+\,km. 
    The wireless access platforms feature 5G-and-beyond MIMO systems operating at the 460--776\,MHz, 3.4--3.6\,GHz, and 27.5--28.35\,GHz bands and with 14, 192, and 384 antenna elements per sector respectively, and they offer up to 3+\,Gbps wireless access throughput and up to 10+\,km effective cell radius. 
With both software-defined radios and programmable COTS systems, and through effective orchestration of these wireless resources with fiber as well as compute resources embedded end-to-end across user equipment (UE), base stations (BS), edge, and cloud, including support for Bring Your Own Device (BYOD), \ara\ offers programmability, performance, robustness, and heterogeneity at the same time, thus enabling rural-focused co-evolution of wireless and applications while helping advance the frontiers of wireless systems in domains such as Open~RAN, NextG, and agriculture applications. The resulting solutions hold the potential of reducing the rural broadband cost by a factor of 10 or more, thus making rural broadband as affordable as urban broadband. 
    Here we present the design principles and implementation strategies of \ara, characterize its performance and heterogeneity, and highlight example wireless and application experiments uniquely enabled by~\ara. 
\end{abstract}




\end{frontmatter}




\section{Introduction}
\label{sec:introduction}

Much like water and electricity, broadband Internet access has become an essential utility. Its significance extends to our professional, educational, and personal lives, as well as to various industries (e.g., precision agriculture) and community services (e.g., public safety). Nonetheless, 39\% of the rural U.S$.$ and 75\% of school-age children 
in rural regions worldwide lack broadband access, most agriculture farms are not connected at all, the current terrestrial broadband technologies are largely not optimized for rural regions, and satellite communications 
incur larger latency (e.g., over 100\,ms even for LEO systems) and offer relatively more limited capacity than the latest terrestrial communication systems such as 5G \cite{ARA:vision, UNICEF-broadband-challenge}. 
    In addition, broadband evolves at a fast pace, and new generations of technologies and services get rolled out every 5--10 years in and around urban regions. Therefore, besides addressing today's rural broadband challenge through mechanisms such as government subsidy, we need to make sure rural regions do not perpetually lag behind urban regions in broadband access, 
    and we need to enable rural-focused wireless technology research and innovation. 

At the same time, rural regions provide rich, diverse use cases of advanced wireless systems, ranging from crop nitrate sensing to 360\degree\ video streaming for rural education, XR-based teleoperation of unmanned ground vehicles (UGVs) and unmanned aerial vehicles (UAVs) for precision farming, collaborative UGVs and UAVs in agriculture automation, and so on. 
In fact, rural regions provide unique real-world opportunities for piloting applications such as UGV and UAV teleoperation to mature emerging wireless technologies and applications in open rural settings before their trials in dense urban environments, thus helping advance the frontiers of wireless systems in general. 

To address the rural broadband challenge and to leverage the rural wireless opportunities, we develop \ara\footnote{Here ARA stands for \underline{A}griculture and Ru\underline{ra}l Communities. In astronomy,
Ara is a southern constellation of stars.} \cite{ARA:vision} as the first wireless living lab for rural-focused research and innovation (R\&I) in advanced wireless systems and their applications in precision agriculture, community services, and so on. To this end, the design and implementation of \ara\ feature the following distinguishing \emph{principles}. 

\subHeading{Rural context.} 
To support rural-focused R\&I, \ara\ captures the unique community, application, and economic context of rural wireless systems through its deployment across a rural area of diameter over 30\,km and including the \ISU\ campus, \ames, \boone, \gilbert, and surrounding research and producer farms as well as rural communities
in \iowa.
In addition, \ara\ features the first-of-its-kind, real-world deployment of long-distance, high-capacity wireless x-haul and access platforms. 
        With wireless x-haul platforms operating at the 11\,GHz, 14\,GHz, 71--86\,GHz, and 194\,THz bands and offering communication capacities up to 160\,Gbps at distances up to 15+\,km, the \ara\ x-haul exemplifies low-cost, high-capacity middle-mile solutions connecting remote rural communities and agriculture farms to the nearest wired Internet backbone. 
    With 5G-and-beyond Multiple-Input Multiple-Output (MIMO) systems operating at the 460--776~\,MHz, 3.4--3.6\,GHz, and 27.5--28.35\,GHz bands and with 14, 192, and 384 antenna elements per sector, respectively, the \ara\ radio access network (RAN) exemplifies massive MIMO wireless access platforms that offer high 
    capacity and large cell radius through beamforming, thus reducing the required spatial density of RAN base stations~(BSes) and cost of rural wireless. 

\subHeading{Wireless and application co-evolution.} 
With a sharp focus on rural wireless applications and for engaging application communities in the R\&I process, \ara\ supports \emph{programmability and measurability} for both wireless and application R\&I. More specifically, \ara\ embeds open-access compute resources end-to-end across UEs, BSes, edge, and cloud, thus enabling programmability and measurability at the network, transport, and applications layers. 
    For link and physical layers, \ara\ uses software-defined-radios (SDRs) for full programmability and measurability at the UEs and BSes, and it employs commercial-off-the-shelf (COTS) RAN and x-haul platforms that have rich APIs for link and physical layer configuration and measurement, including the first-of-its-kind open API for a television-white-space (TVWS) massive MIMO platform. 
To ensure high \emph{performance and robustness} that are required by application communities but challenging to achieve for early-stage wireless R\&I, \ara\ employs high-performance, mature COTS 5G-and-beyond RAN platforms as well as high-capacity, long-distance wireless x-haul platforms, it deploys high-performance compute resources end-to-end from UEs to the cloud, including the deployment of GPUs at the edge and cloud for applications such as real-time image processing. 
    The design of \ara\ facilitates the use of spatial, temporal, and spectral diversity to improve the robustness of x-haul and RAN communications, and the use of weather sensors for predictive network adaptation. 

\subHeading{Efficient, Trustworthy \& Reproducible Experimentation.} 
For effective resource management and experiment orchestration, we architect \ara\ such that its wireless x-haul and access platforms are managed by their associated compute resources (e.g., host computers). This way, \ara\ can leverage OpenStack \cite{OpenStack} software platform to manage its compute resources and then extend  OpenStack to manage \ara's wireless x-haul and access resources, for instance, orchestrating resource allocation to avoid wireless interference between concurrent experiments and safe-guarding \ara\ from misbehaving wireless experiments that may violate wireless spectrum use specification. 
    The use of widely-accepted software platforms such as OpenStack and Docker~\cite{Docker} in \ara\ helps minimize the learning curves of \ara\ users, it facilitates the potential translation of \ara-enabled research results 
    into industry practice (e.g., in the OpenStack ecosystem), and it enables ready integration of \ara\ with other national cyberinfrastructures such as the programmable network backbone FABRIC \cite{FABRIC} and the research cloud Chameleon \cite{NSFCloud:Chameleon}. Further, \ara\ provides a programmable interface for specifying and executing experiments via Jupyter Notebooks~\cite{jupyter}, allowing users to specify, share, and reproduce experiments in a rigorous and effective manner, 
    thereby enabling reproducibility and community collaboration in \ara-enabled research. 

Guided by the above design and implementation principles, with the first-of-its-kind deployment of advanced wireless and computing platforms in real-world agriculture and rural settings, and through effective resource management and experiment orchestration including support for Bring Your Own Device (BYOD) experiments, \ara\ enables unique R\&I experiments in a wide range of topics. As we will discuss in detail in Section~\ref{sec:ara-enabled-research}, ARA-enabled R\&I spans O-RAN, open-source NextG, TVWS massive MIMO, non-terrestrial networks~(NTN), spectrum innovation, integrated rural wireless access and x-haul, 360\degree\ video streaming, real-time edge data analytics, agriculture automation, and so on.  
    The resulting solutions hold the potential of reducing the rural broadband cost by a factor of 10 or more, thus making rural broadband as affordable as urban broadband today \cite{ARA:vision}. 

In what follows, we elaborate on the \ara\ design and implementation in Sections~\ref{sec:design} and \ref{sec:implementation}, respectively. We evaluate the performance and heterogeneity  
of \ara, and we present exemplars of \ara-enabled wireless and application experiments in Section~\ref{sec:EvalExperiments}. We discuss related work in Section~\ref{sec:related-work}, and make concluding remarks in Section~\ref{sec:concludingRemarks}.


\section{Design of \ara} 
\label{sec:design}

We now delve into
\ara\ infrastructure and software designs to capture rural community, application, and economic context, to support wireless and application co-evolution, and to effectively manage \ara\ resources and orchestrate research experiments.

\subsection{\ara\ Deployment for Capturing Real-World Rural Context} 
\label{subsec:ara-deployment}

Rural communities tend to be sparsely distributed, with inter-community distances being tens of kilometers or more. These communities, together with surrounding agriculture farms, are usually far away from the wired Internet backbone. The generally sparse user spatial distribution makes it not economically viable to pervasively deploy in rural regions urban-focused broadband technologies such as fiber and small cells; we need affordable \emph{middle-mile} solutions to connect rural communities and agriculture farms with one another and the Internet, and affordable \emph{last-mile} solutions to connect end-user devices such as automated ground and aerial vehicles in precision agriculture \cite{ARA:vision}. 
    This unique community and economic context motivates the use of two broad categories of wireless solutions in rural regions: high-capacity, long-distance terrestrial and non-terrestrial wireless x-haul networks as middle-mile solutions and high-capacity wireless access networks as last-mile solutions. 

\begin{figure*}[!htbp]
  \centering
\includegraphics[width=\textwidth]{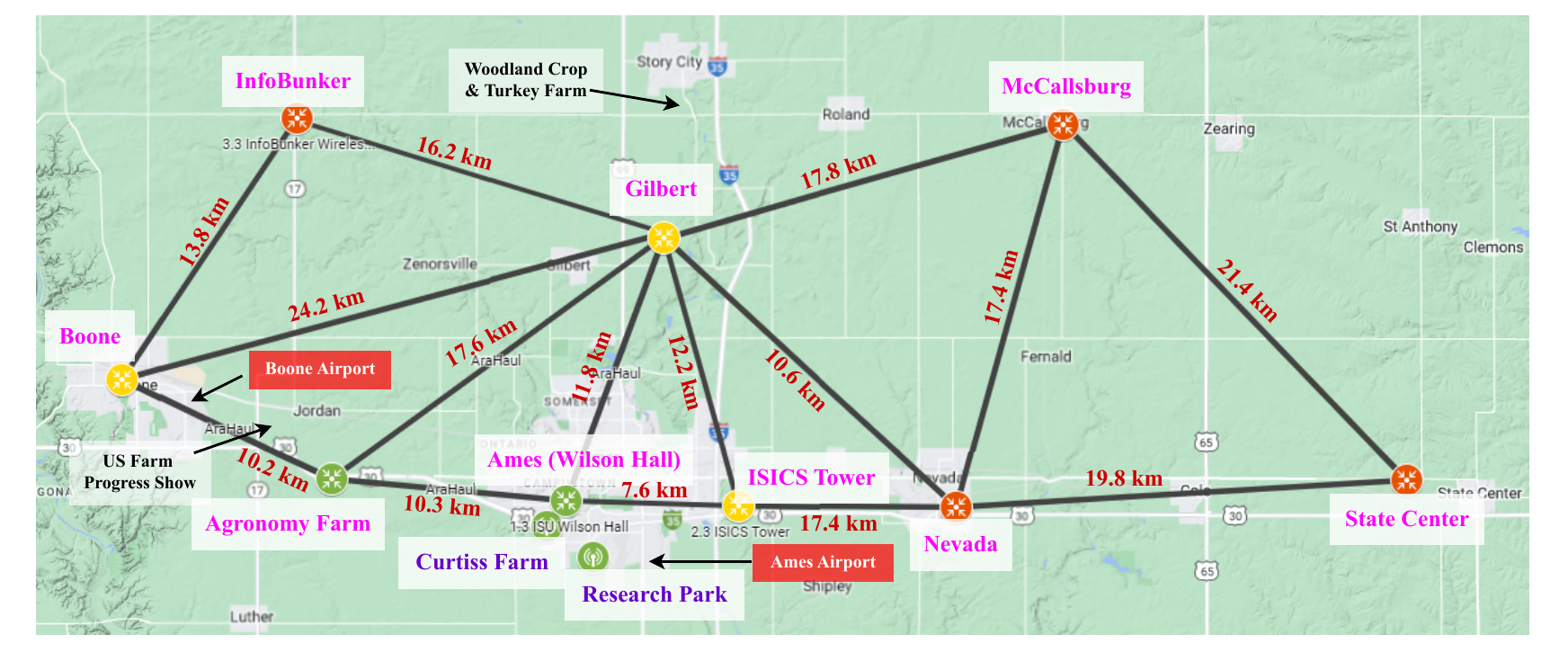}
  \caption{\ara\ deployment  
  }
  \label{fig:ara-deployment}
\end{figure*}

Accordingly, we design \ara\ so that its deployment covers a rural area of diameter over 30\,km and includes \ISU\ campus, \ames, \boone\ and \gilbert, and surrounding research and producer farms as well as rural communities in \iowa, 
as shown in Figure~\ref{fig:ara-deployment}. 
\ara\ consists of \arahaul\ and \araran. \arahaul\ is a high-capacity wireless x-haul mesh network spanning \ames, \agronomyfarm, \boone, and \gilbert. 
\araran\ is a high-capacity Radio Access Network (RAN) with Base Stations (BSes) deployed at every \arahaul\ site as well as \curtissfarm\ and \researchpark. 
\infobunker\ site will be deployed in 2025 while \nevada, \McCallsburg,  and \statecenter\ sites will be deployed in \mbox{Phase-3} of \ara. The deployment of \ara\ around the \ISU\ campus and the \researchpark\ that host corporate research centers of leading AgTech and avionics companies such as John Deere, Vermeer, and Collins Aerospace helps engage industry in \ara-enabled application innovation, which in turn drives innovation in advanced wireless.

To capture the application context of rural wireless, \ara\ deploys 50+ User Equipment (UEs) across agriculture farms and rural communities around the \araran\ 
sites.
The UE deployment features both crop and livestock farms, farm facilities (e.g., grain mills, grain bins, and biorefineries), as well as agriculture vehicles, robots, UAVs, and phenotyping cameras and sensors. In rural communities, the UEs are deployed at city water towers, municipal airports, buses, and public safety vehicles such as fire and police vehicles. 
    The UE deployment captures diverse rural wireless use cases in agriculture and communities, and it enables integrative wireless and application experiments in real-world rural settings, thus facilitating wireless and application co-evolution.

\subsection{\ara\ Architecture and Platforms for Wireless and Application Co-Evolution} 
\label{subsec:co-evolution-design}

With \ara\ deployment capturing the unique rural community and application context, we architect and provision \ara\ with carefully-selected wireless and compute platforms, weather and RF sensors, as well as smart Power Distribution Units (PDUs) to enable integrative wireless and application experiments.

\subsubsection{\ara\ System Architecture}
\label{subsubsec:ara-architecture}

\begin{figure*}[t!]
    \centering
    \hspace{-.5cm}
    \includegraphics[width=\textwidth]{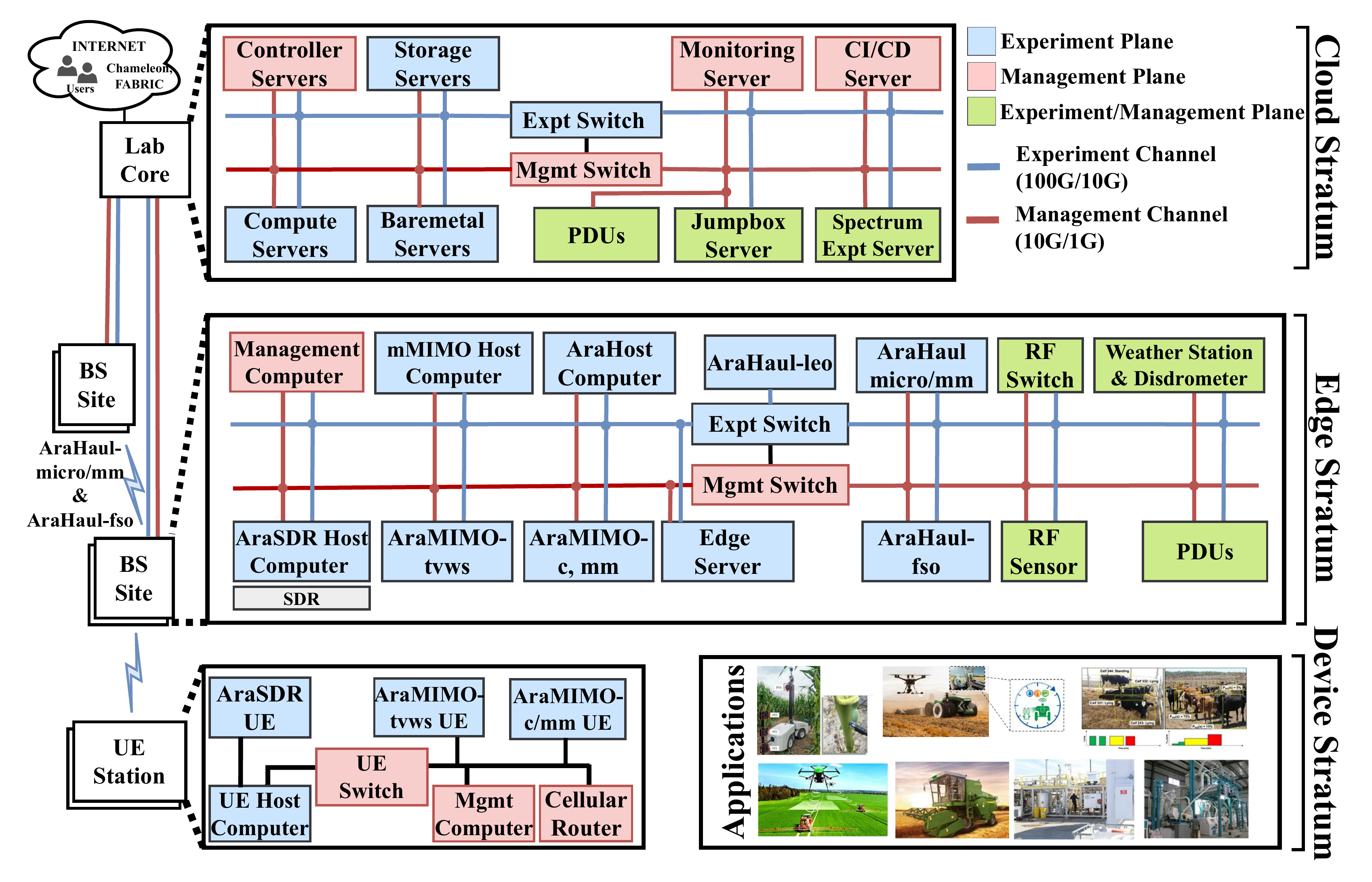}
    \caption{\ara\ system architecture 
    }
    \label{fig:system-architecture}
\end{figure*}

To enable high-fidelity wireless and applications experiments, \ara\ is architected to reflect real-world architectures of rural wireless systems, as shown in Figure~\ref{fig:system-architecture}.

More specifically, the \ara\ infrastructure consists of the \emph{device stratum, edge stratum}, and \emph{cloud stratum}, reflecting the device-edge-cloud continuum in today's wireless and application infrastructures. 
    The \emph{device stratum} includes the \ara\ UE stations together with application-specific equipment and their associated sensors and actuators (e.g., agriculture vehicles of leading AgTech companies like John Deere and Vermeer, with high-resolution imaging sensors and weed-sprayer-nozzles). 
The \emph{edge stratum} includes the \ara\ BS sites and the networks (e.g., \arahaul) connecting these BS sites to one another. The edge stratum provides wireless connectivity to the \ara\ UE stations, it provisions edge computing resources for processing time-sensitive data from the device stratum, and it connects the device stratum to the cloud stratum for enabling cloud resource access by rural applications such as agriculture automation. 
    The \emph{cloud stratum} includes the \ara\ lab core where the \aracontroller\ as well as wireless- and application-oriented cloud compute, storage, and network resources are provisioned. As we will discuss in more detail in Section~\ref{subsec:arasoft-design}, the \aracontroller\ manages \ara\ resources and orchestrates experiments in \ara. 
Through the National Science Foundation (NSF) FABRIC network \cite{FABRIC} and Internet, \ara\ is connected to other experimental infrastructures such as the Chameleon cloud \cite{NSFCloud:Chameleon}. 

To enable end-to-end orchestration of the resources from the device, edge, and cloud strata for 
experiments, the resources at the \ara\ UE stations, BS sites, and lab core are organized into the experiment plane and management plane, supporting wireless and application experimentation and \ara\ management, respectively. 
    In the \emph{experiment plane}, both the UE stations and BS sites have the rural-focused \araran\ wireless platforms, most BS sites have rural-focused \arahaul\ wireless platforms, a subset of the BS sites have edge compute resources, and the lab core has cloud compute and network resources. All the BS sites and the lab core are connected through dedicated fiber networks and/or \arahaul\ wireless platforms. All the BS sites and UE stations have RF sensors, and a subset of the BS sites also have weather sensors; these \emph{RF and weather sensors} enable spectrum innovation research and the characterization of weather impact 
    on rural wireless communications. The UE stations, BS sites, and lab core have \emph{smart Power Distribution Units (PDUs)} to enable the power consumption measurement and on/off control of the individual equipment, thus enabling research in energy-efficient wireless systems and applications. 
In the \emph{management plane}, the lab core has dedicated servers for controlling \ara\ resource management and experiment orchestration, monitoring \ara\ health status and spectrum use integrity, and managing the continuous evolution of the \ara\ software systems and associated Continuous Integration and Continuous Deployment (CI/CD) processes. 
    The BS sites and UE stations have dedicated management computers that interact with the lab core in resource management and experiment orchestration. The BS sites connect to the lab core through a dedicated management network of fibers and wireless backhauls, and the UE stations connect to the lab core through management paths provided by commercial cellular networks. 
The data traffic in the experiment plane and management plane is isolated from each other since each plane has its own dedicated network or VLAN. 
    The UE stations, BS sites, and lab core are architected so that new devices can be easily integrated into the experiment plane and management plane, thus facilitating BYOD experiments.

\subsubsection{\ara\ Experiment Platforms} %

To support \emph{diverse rural use cases}, \ara\ features heterogeneous wireless access and x-haul platforms, as shown in Table~\ref{tab:araran-spectrum}. 
\begin{table}[ht!] {
    \centering
    \caption{Heterogeneous, High-Capacity Wireless Access and x-haul Platforms in \ara  
    }
    {\small
    \resizebox{\textwidth}{!}
    {
    \begin{tabular}{| p{.05in} |l | l | l | l | l |}
    \cline{2-6}
    \multicolumn{1}{c | }{}& \makecell[l]{ \textbf{Platform}} & \makecell[l]{\textbf{Frequency}} & \makecell[l]{\textbf{Bandwidth}} &
    \makecell[l]{\textbf{Capacity}} &
    \makecell[l]{\textbf{Range}}
    \\  \cline{2-6}
    \multicolumn{4}{c }{}\\[-2.3ex]
    \hline
    \multirow{4}{*}{\rotatebox{90}{\textbf{AraRAN}}}
    & \aratvws\ & 460--776\,MHz & upto 40\,MHz & 100+\,Mbps & 8.5+\,km \\ \cline{2-6} 
    & \arac\ & 3.45--3.55\,GHz & 100\,MHz & 650+\,Mbps & 8.5+\,km \\\cline{2-6}
    & \aramm\ & 27.5--27.9\,GHz & $4\times100$\,MHz & 1.3+\,Gbps & 500+\,m \\ \cline{2-6}
    & \arasdr\ & 3.4--3.6\,GHz & 200\,MHz & 100+\,Mbps & 1.2+\,km \\ \cline{2-6}  
    \hline
    \multirow{3}{*}{\rotatebox{90}{\textbf{AraHaul}}} 
    & \aviatmicro\ &   10.6--11.5\,GHz  & 100\,MHz & 1\,Gbps & 20+\,km \\ \cline{2-6}  
    & \aviatmm\ & 71--86\,GHz & 2\,GHz & 10\,Gbps & 15+\,km \\ \cline{2-6}  
    & \araoptical\ & 191.7–194.8\,THz & 80\,GHz & 160\,Gbps & 10+\,km \\ \cline{2-6}  
    & \arantn\ & 10--14.5\,GHz & 250\,MHz & 195\,Mbps & Global \\ \hline  
    \end{tabular}
    }
    \label{tab:araran-spectrum}
    }
}
\end{table}
Despite the generally sparse spatial distribution of users, the UE spatial density in rural cities and communities as well as during the busy periods of agriculture farms (e.g., the weeks of harvesting season in early fall) can be close to that in suburban or even urban regions. Therefore, \araran\ adopts massive MIMO wireless access platforms spanning the low-band, mid-band, and high-band, and they exemplify wireless systems that offer high capacity and large cell radius through beamforming, thus reducing the required spatial density of RAN base stations and the cost of rural wireless. 
    Similarly, \arahaul\ leverages diverse long-distance, high-capacity x-haul platforms; it employs the 11\,GHz and 71--86\,GHz x-haul COTS platforms from Aviat Networks, denoted by \aviatmicro\ and \aviatmm\ respectively, and it features the first-of-its-kind long-distance terrestrial free-space optical communications platform \araoptical\ which we develop for \ara\  (see Section~\ref{subsec:araoptical}). 
    In addition to the terrestrial x-haul links, \ara\ also features a Low Earth Orbit~(LEO) satellite backhaul link, i.e., \arantn, deployed at Wilson hall with Indoor Unit~(IDU) from Hughes which connects to the OneWeb satellites.
These \arahaul\ platforms exemplify low-cost, high-capacity middle-mile solutions for rural regions. 

\begin{table}[]
\caption{\ara\ Infrastructure and Software Platforms}
\label{tab:ara-resources}
\resizebox{1.1\textwidth}{!}{
\begin{tabular}{cc|l|l|}
\cline{3-4}
\multicolumn{2}{l|}{}                                                                                                                                         & \multicolumn{1}{c|}{\textbf{Component}}                                                                             & \multicolumn{1}{c|}{\textbf{Specification}}                                                                                           \\ \hline
\multicolumn{1}{|c|}{\multirow{29}{*}{\rotatebox{90}{\textbf{\sc  Infrastructure}}}} & \multirow{4}{*}{\textbf{Lab Core}}                                                      & \begin{tabular}[c]{@{}l@{}}Controller, Storage, Compute, Baremetal\\ Servers, CI/CD, and Monitoring Nodes\end{tabular} & \begin{tabular}[c]{@{}l@{}}Dell PowerEdge R750/R760, Intel Xeon CPU, \\ 24--48 Cores, 64--384\,GB RAM, 1G/10G/100G NIC\end{tabular} \\ \cline{3-4} 
\multicolumn{1}{|c|}{}                                              &                                                                                         & Jumpbox Server                                                                                                      & \begin{tabular}[c]{@{}l@{}}SuperMicro, AMD Opteron 6276, 16 Cores, \\ 32\,GB RAM, 10G NIC\end{tabular}                                 \\ \cline{3-4} 
\multicolumn{1}{|c|}{}                                              &                                                                                         & Management Switch                                                                                                   & Cisco Catalyst 9300                                                                                                                   \\ \cline{3-4} 
\multicolumn{1}{|c|}{}                                              &                                                                                         & Experiment Switch                                                                                                   & Juniper ACX7100                                                                                                                       \\ \cline{2-4} 
\multicolumn{1}{|c|}{}                                              & \multirow{15}{*}{\textbf{\begin{tabular}[c]{@{}c@{}}Base Station \\ Site\end{tabular}}} & \makecell[l]{SDR Host, x-haul Host, Management\\ Servers, and Edge Servers}                                                                      & \begin{tabular}[c]{@{}l@{}}Dell PowerEdge R750, Intel Xeon CPU, \\ 8--24 Cores, 64\,GB RAM, 1G/10G NIC\end{tabular}              \\ \cline{3-4} 
\multicolumn{1}{|c|}{}                                              &                                                                                         & AraMIMO-tvws                                                                                                        & Skylark Wireless Faros V2 CU, DU, RU                                                                                                  \\ \cline{3-4} 
\multicolumn{1}{|c|}{}                                              &                                                                                         & AraMIMO-c                                                                                                           & Ericsson Air 6419 RU and Baseband 6647 gNB                                                                                                                    \\ \cline{3-4} 
\multicolumn{1}{|c|}{}                                              &                                                                                         & AraMIMO-mm                                                                                                          & Ericsson Air 5322 RU and Baseband 6647 gNB                                                                                                                    \\ \cline{3-4} 
\multicolumn{1}{|c|}{}                                              &                                                                                         & AraSDR                                                                                                              & USRP N320, CommScope Antennas, Tower Mounted Boosters                                                                                                    \\ \cline{3-4} 
\multicolumn{1}{|c|}{}                                              &                                                                                         & AraHaul-micro and AraHaul-mm                                                                                           & \makecell[l]{Aviat WTM 4811 (11G+18G), 4200 (11G), \\
MBXD (4800 for 80G and 4200 for 11G)}                                                                                                                         \\ \cline{3-4} 
\multicolumn{1}{|c|}{}                                              &                                                                                         & AraHaul-fso                                                                                                         & Custom-made FSOC Telescope                                                                                                            \\ \cline{3-4} 
\multicolumn{1}{|c|}{}                                              &                                                                                         & AraHaul-leo                                                                                                         & Hughes IDU and Satellite Terminal                                                                                                     \\ \cline{3-4} 
\multicolumn{1}{|c|}{}                                              &                                                                                         & RF Sensor                                                                                                           & Keysight N6841A                                                                                                                             \\ \cline{3-4} 
\multicolumn{1}{|c|}{}                                              &                                                                                         & RF Switch                                                                                                           & Keysight RF Switch                                                                                                                             \\ \cline{3-4} 
\multicolumn{1}{|c|}{}                                              &                                                                                         & Weather Sensor                                                                                                      & Davis Vantage Pro2                                                                                                                   \\ \cline{3-4} 
\multicolumn{1}{|c|}{}                                              &                                                                                         & Disdrometer                                                                                                         & OTT Parsivel\textsuperscript{2}                                                                                                                         \\ \cline{3-4} 
\multicolumn{1}{|c|}{}                                              &                                                                                         & Smart PDUs                                                                                                          & CyberPower PDU81001, PDU81003                                                                                                                           \\ \cline{3-4} 
\multicolumn{1}{|c|}{}                                              &                                                                                         & Management Switch                                                                                                   & Cisco Catalyst 9300                                                                                                                   \\ \cline{3-4} 
\multicolumn{1}{|c|}{}                                              &                                                                                         & Experiment Switch                                                                                                   & Juniper ACX710                                                                                                                        \\ \cline{2-4} 
\multicolumn{1}{|c|}{}                                              & \multirow{8}{*}{\textbf{UE Station}}                                                    & UE Host Computer                                                                                                    & \begin{tabular}[c]{@{}l@{}}SuperMicro E300 Miniserver, Intel Xeon D-1736NT, \\ 8~Cores, 32\,GB RAM, 1G/25G Interfaces\end{tabular}     \\ \cline{3-4} 
\multicolumn{1}{|c|}{}                                              &                                                                                         & UE Management Computer                                                                                              & \begin{tabular}[c]{@{}l@{}}Dell Optiplex 3090, Intel Core i5, \\ 6 Cores, 8 GB RAM, 1G Interface\end{tabular}                         \\ \cline{3-4} 
\multicolumn{1}{|c|}{}                                              &                                                                                         & AraMIMO-tvws UE Radio                                                                                               & Skylark Faros V2 CPE with Cross-polarized Directional Antenna                                                                                                               \\ \cline{3-4} 
\multicolumn{1}{|c|}{}                                              &                                                                                         & AraMIMO-c/mm UE Radio                                                                                               & Quectel RG530                                                                                                                         \\ \cline{3-4} 
\multicolumn{1}{|c|}{}                                              &                                                                                         & Spectrum Monitor SDR                                                                                                & USRP B205                                                                                                                             \\ \cline{3-4} 
\multicolumn{1}{|c|}{}                                              &                                                                                         & AraSDR UE Radio                                                                                                     & USRP B210 with UE Booster Amplifier and Omni Antenna                                                                                                                \\ \cline{3-4} 
\multicolumn{1}{|c|}{}                                              &                                                                                         & UE Switch                                                                                                           & Juniper EX2300                                                                                                                        \\ \cline{3-4} 
\multicolumn{1}{|c|}{}                                              &                                                                                         & Cellular Router                                                                                                     & Cradlepoint IBR600C                                                                                                                   \\ \cline{2-4} 
\multicolumn{1}{|c|}{}                                              & \multirow{2}{*}{\textbf{\begin{tabular}[c]{@{}c@{}}Measurement \\ Tools\end{tabular}}}  & Field Measurements and Testing                                                                                      & NEMO UE, One Plus UE                                                                                                                  \\ \cline{3-4} 
\multicolumn{1}{|c|}{}                                              &                                                                                         & Spectrum Sensing                                                                                                    & Keysight Fieldfox                                                                                                                     \\ \hline
\multicolumn{1}{|c|}{\multirow{9}{*}{\rotatebox{90}{\textbf{\sc Software}}}}  & \multirow{5}{*}{\textbf{Platform}}                                                & Cloud Operating System (AraController)                                                                              & OpenStack Yoga                                                                                                                        \\ \cline{3-4} 
\multicolumn{1}{|c|}{}                                              &                                                                                         & Server Operating System, Container Engine                                                                                                    & Ubuntu 20.04/22.04 LTS, Docker                                                                                                              \\ \cline{3-4} 
\multicolumn{1}{|c|}{}                                              &                                                                                         & Network Monitoring/Health Statistics                                                                              & LibreNMS, Prometheus, Custom APIs                                                                                                     \\ \cline{3-4} 
\multicolumn{1}{|c|}{}                                              &                                                                                         & Time Synchronization                                                                                                           & Network Time Protocol~(NTP), Precision Time Protocol~(PTP)                                                                                                                              \\ \cline{3-4} 
\multicolumn{1}{|c|}{}                                              &                                                                                         & Security                                                                                                            & \texttt{auditd}, Linux \texttt{iptables}                                                                                                             \\ \cline{2-4} 

\multicolumn{1}{|c|}{}                                              & \multirow{2}{*}{\textbf{AraRAN}} & BS and UE                                                                                                                 & srsRAN, OpenAirInterface~(OAI)                                                                                                                           \\ \cline{3-4} 
\multicolumn{1}{|c|}{}                                              &                                                                                         & Core Network                                                                                                               & Open5GS, OAI Core, Aether SD-Core                                                                                                     \\ \cline{2-4}
\multicolumn{1}{|c|}{}                                              & \textbf{\aratvws} & BS and UE                                                                                                                 & Platform-specific software, Custom APIs                                                                                                                           \\ \cline{2-4} 
\multicolumn{1}{|c|}{}                                              & \textbf{AraMIMO-c/mm} & BS and UE                                                                                                                 & Platform-specific software, Moshell, Custom APIs                                                                                                                           \\ \cline{2-4} 
\multicolumn{1}{|c|}{}                                              & \textbf{AraHaul} & x-haul Radios and FSOC Telescopes                                                                                                                & Platform-specific firmware, Custom APIs                                                                                                                           \\ \cline{2-4}

\multicolumn{1}{|c|}{}                                              & \textbf{Measurement} & COTS AraRAN/AraHaul                                                                                                                 & Custom APIs                                                                                                                           \\ \cline{2-4}
\multicolumn{1}{|c|}{}                                              & \textbf{Application}                                                                    & Video Streaming                                                                                                     & GStreamer, QUIC, QUICHE                                                                                                               \\ 
\hline

\end{tabular}}
\end{table}

Regarding \araran, the \emph{low-band} \aratvws\ platform utilizes the COTS Skylark Faros 
massive MIMO system at the TVWS band; with 14~antenna elements per sector, it provides a cell coverage up to 10+\,km and per-user-group communication capacity up to 100+\,Mbps at the same time, thus suitable for connecting UEs in the country side and far away from the base stations. 
    The \emph{mid-band} \mbox{\arac\ }platform utilizes the COTS Ericsson AIR 6419 
    massive MIMO system operating at the 3.4--3.6\,GHz band; with 192~antenna elements per sector, it provides a cell coverage up to 8.5+\,km and per-user-group communication capacity up to 650+\,Mbps, thus suitable for connecting general UEs in the city/community center and agriculture farms. 
The \emph{high-band} \aramm\ platform utilizes the COTS Ericsson AIR 5322 
mmWave system operating at the 27.5--28.35\,GHz band; with 384~antenna elements per sector, it provides a cell coverage up to 500+\,meters and per-user-group communication capacity up to 3+\,Gbps, thus suitable for dense city centers and during the busy periods of agriculture farm operations. 
    The mid-band portion of \araran\ also uses USRP SDRs N320 and B210 
    to support open-source NextG and O-RAN experiments. 

For rural-focused wireless and application R\&I, \araran\ and \arahaul, together with the \ara\ platforms for compute, weather and RF sensing, as well as power metering and control, support \emph{whole-stack programmability and measurability}. 
    As shown in Figure~\ref{fig:system-architecture}, open-access compute resources are embedded end-to-end across UE stations, BS sites, and lab core. Together with open-source software systems such as the Ubuntu Linux operating system and softwarized telecom and application platforms (e.g., Open5GS, DPDK, QUICHE, and GStreamer), 
    these compute resources enable programmability and measurability at the \emph{network, transport, and application layers} to support experiments at these layers across all \ara\ wireless and compute platforms. 
For programmability and measurability at the \emph{link and physical layers}, the USRP SDRs N320 and B210, together with open-source 5G software platforms such as OpenAirInterface~(OAI)~\cite{oai5g:online}, srsRAN~\cite{srsRAN:online}, and O-RAN Software Community systems, 
enable full programmability at the \araran\ BS sites and UE stations. 
    Through effective \emph{software wrappers} (as we will elaborate in Section~\ref{subsec:COTS-programmability}), \ara\ also provides rich APIs for configuring and measuring the link and physical layers of its COTS wireless access and x-haul platforms, including the first-of-its-kind open API for the TVWS massive MIMO platform \aratvws. 
\ara\ also uses similar software wrapper approaches to enable the configuration and measurement with weather and RF sensors, as well as smart PDUs. 
Therefore,
    \ara\ enables comprehensive measurement of wireless channel and link behavior (e.g., channel path loss, massive MIMO channel state matrix, and link bit error rate), weather condition (e.g., rain rate 
    and raindrop 
    size), and equipment power consumption, and it enables configuration of important link and physical layer behavior of wireless platforms (e.g., massive MIMO user grouping and beamforming, operating frequency, transmission power, as well as modulation and coding schemes).

The design of \ara\ also pays special attention to the \emph{performance and robustness} required by wireless and applications R\&I. 
    For instance, application R\&I tend to require performance and robustness that are challenging to achieve for early-stage wireless R\&I and commonly used open-source 5G platforms (e.g., OpenAirInterface), whose performance lags behind the commercial wireless platforms as of today, e.g., in terms of capacity and reliability in outdoor real-world settings that is often required by rural applications. \ara\ addresses the challenge by employing high-performance, mature COTS 5G-and-beyond RAN platforms as well as high-capacity, long-distance wireless x-haul platforms while enabling whole-stack programmability as mentioned earlier.  
Together with the diverse operation frequencies of \araran\ and \arahaul\ platforms, the mesh network topology and the specific designs of the \ara\ BS sites and UE stations (see Sections~\ref{subsec:BS-site} and \ref{subsec:UE-station}) also facilitate the use of spatial, temporal, and spectral diversity as well as weather-sensor-enabled predictive network adaptation to improve the robustness of x-haul and RAN communications in inclement weathers. 

The \ara\ compute and storage resources have also been carefully designed to support the needs of wireless and application R\&I. 
For instance, the compute and baremetal servers at the Cloud Stratum have high-performance specs (e.g., 3\,GHz and 48 Core Intel Xeon 6326 CPU, 384\,GB RAM) for functions such as those of 5G core networks and real-time data processing. 
    The storage servers are equipped with 28\,TB storage each and in RAID-5 configuration to ensure fault tolerance. 
    At the Edge Stratum, each SDR host computer supports three SDRs and has high-performance specs (3.1\,GHz, 16 Core Intel Xeon 6346 CPU, 64\,GB RAM) for open-source 5G signal processing and high-layer network stacks. 
    The edge stratum also hosts edge servers with NVIDIA RTX~5000 GPUs 
    for real-time processing of application data such as the multi-spectral, high-resolution imaging data used in agriculture automation \cite{RT-LWN}.
At the Device Stratum, the UE host computers (with 2.7\,GHz, 8 Core Intel Xeon D-1736NT CPU, 32\,GB RAM) support experiments with UE SDRs as well as 
\aratvws, \arac, and \aramm\ Radios. 


\vspace{-.1in}
\subsection{Efficient, Trustworthy and Reproducible Experimentation 
}
\label{subsec:arasoft-design}

To manage the heterogeneous \ara ~resources 
and orchestrate diverse user experiments with maximum automation, we develop the \ara\ software framework \arasoft. \arasoft\ has the following key features:
\begin{itemize}[leftmargin=*]
   \item {\bf Efficiency.} \arasoft\ leverages OpenStack~\cite{OpenStack} to achieve efficient resource management and provisioning. For instance, it takes less than 15 seconds for the ARA Resource Manager to create a resource lease on a user request. 
   Moreover, we employ a set of best practices (e.g., CI/CD~\cite{WhatisCI53:online}) to ensure that \ara~ is efficient and sustainable in terms of 
   long-term maintenance and evolution. 

    \item {\bf Trustworthiness.} 
    \arasoft\ includes fine-grained health monitoring, fault tolerance, 
    and security mechanisms to ensure high availability, reliability, and security of \ara\ experiments. For instance, $\sim$80\% of the failures of the COTS wireless links can recover within 30~seconds. 

    \item {\bf Reproducibility.} In addition to an intuitive graphical user interface, 
    \arasoft\ provides a programmable interface for specifying and executing experiments via Jupyter Notebooks~\cite{jupyter}, allowing users to specify, share, and reproduce experiments in a rigorous and effective manner. 
    Moreover, it includes a rich set of APIs to enable the customization of heterogeneous wireless platforms for reproducible experiments.

\end{itemize}

We elaborate on a few key design tradeoffs of \arasoft\ below. 
First, unlike many other  testbeds~\cite{grid5000,cloudlab,marojevic2020aerpaw,Powder,netbed} which are based on in-house software, we develop \arasoft\ based on the foundation of OpenStack, a proven open-source cloud platform in production. 
This design decision brings multiple benefits,  including leveraging the collective wisdom and experiences from the broad open-source community, 
specifically Chameleon Cloud \cite{chameleon-lessons, chi-in-a-box},
and contributing back to the community to co-shape the future.
In addition, the rich set of cloud services in OpenStack, including user and resource management, networking, and orchestration, make it a best-fit baseline for \arasoft.

\begin{figure}[t!]
    \centering
    \includegraphics[width=\textwidth]{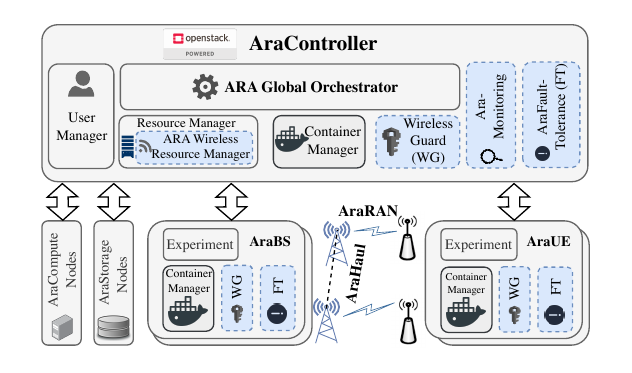}
    \caption{AraSoft overview (Blue dash-lined boxes denote the components newly introduced by AraSoft atop OpenStack).
    }
    \label{fig:arasoft_top_view}
\end{figure}

With the solid foundation, we further introduce a set of unique functionalities to support the  \ara\ vision. As shown in Figure~\ref{fig:arasoft_top_view}, \arasoft\ consists of  1) AraController for overall user management and experiment control, 2) AraBS and AraUE for BS and UE management, 
and 3)  AraCompute and AraStorage for compute and storage management, respectively.
The blue dash-lined boxes highlight the new 
components that \arasoft\ introduces to the OpenStack ecosystem.

More specifically, AraController serves as the operating system for \ara.
It includes a \textit{User Manager} for user authentication and access control (e.g., federated login via Globus~\cite{globus}), a  \textit{Resource Manager} for resource pool 
management, and a \textit{Container Manager} for executing containerized experiments. 
Moreover, there is an \textit{ARA Global Orchestrator} to coordinate all the activities with dedicated security  and reliability support (e.g., {LibreNMS}~\cite{librenms} and {Prometheus}~\cite{prometheus} based \textit{AraMonitoring}, redundancy-based \textit{AraFaultTolerance}.

One of the most important capabilities of AraSoft is \textit{Wireless Guard} (WG)~\cite{ARA-Wireless-Guard}. In \ara, concurrent users may share resources that utilize the wireless spectrum across various bands. To ensure necessary isolation and  compliance with spectrum usage policies, the WG modules in the AraBS and AraUE continuously monitor users' spectrum usage behavior and coordinate with the WG module in the AraController. WG operates in both the \textit{proactive} and \textit{reactive} modes. In proactive mode, user behavior---such as user-configured SDR parameters including transmission frequency and power---is intercepted from the USRP Hardware Driver~(UHD). On the other hand, the reactive mode adopts a hardware-based approach, continuously monitoring transmitted signals using RF sensors connected to SDRs via directional couplers at the BS (see Figure~\ref{fig:bs-rf}) and UE~(see B205 monitor in Figure~\ref{fig:ue-deployment}). The WG module compares the user behavior from the proactive and reactive modules with the experiment specification and spectrum policy. If it detects any violation of the usage policy or experiment specification, WG immediately terminates the offending experiment. Besides the proactive and reactive approaches, WG enhances security through strict access controls in \ara\ experiment APIs.

\subsection{\ara-Enabled Research}
\label{sec:ara-enabled-research}

Guided by rural-focused wireless living lab design principles, with first-of-its-kind deployment of advanced wireless and computing platforms in real-world agriculture and rural settings, and through effective resource management and experiment orchestration, \ara\ enables unique research and innovation 
ranging from the modeling to the architectures, technologies, services, and applications of rural wireless systems, as shown in Table~\ref{tab:ara_enabled_research}. 
They cover a wide range of topics such as those in Open~RAN, open-source NextG, TVWS massive MIMO, NTN, spectrum innovation, integrated rural wireless access and x-haul, 360\degree\ video streaming, real-time edge data analytics, and agriculture automation.
In Section~\ref{sec:EvalExperiments}, we will characterize the capacity, coverage, and heterogeneity of \ara,  
and we will delve into example wireless and application experiments uniquely enabled by \ara.
\begin{table}[!htb]
    \centering
    \caption{\ara-Enabled Research and Innovation 
    }
    \label{tab:ara_enabled_research}
    {
    \resizebox{\columnwidth}{!}
    {
    \begin{tabular}{|c|l|}
    \hline 
    \makecell[c]{\textbf{Areas}} & \makecell[c]{\textbf{Exemplars}}\\\hline\hline
    
         \makecell[c]{Modeling} &  \makecell[l]
         {$\circ$ Real-world rural wireless channel characterization\\ $\circ$ Real-world characterization of physical dynamics and \\ \hspace*{0.07in} mobility of agriculture UAVs and UGVs \\ } \\\hline
         
         \begin{tabular}{c}
         Network\\Architecture\end{tabular} & \makecell[l]{
         $\circ$ O-RAN architecture for real-time cyber-physical \\ \hspace*{0.07in} systems of agriculture vehicles and robots \\
         $\circ$ Multi-modal, long-distance, and high-throughput\\ \hspace*{0.07in} terrestrial and non-terrestrial wireless x-haul networking\\
         $\circ$ Integrated rural wireless access and x-haul networking\\
         $\circ$ Integrated wireless networking and edge computing}\\\hline
         
         \begin{tabular}{c}
         \makecell[c]{Technology\\
         \& \\ Service}\end{tabular} & \makecell[l]{$\circ$ Ultra-reliable, low-latency communications (URLLC)\\
         $\circ$ Massive MIMO, beam-forming, and beam tracking\\
         $\circ$ Dynamic spectrum sharing\\
         $\circ$ Open-source NextG for rural green networking}\\\hline
         
         \makecell[c]{Application} & \makecell[l]{
         $\circ$ 360\degree~video streaming for agriculture education \\ 
         $\circ$ Real-time video streaming and analytics for agriculture \\ 
         \hspace*{0.07in} automation and livestock health monitoring \\ 
         $\circ$ XR-based teleoperation of agriculture UAVs 
         }
         \\\hline
    \end{tabular}
    }}    
\end{table}

\subsection{\ara\ Accessibility}
\label{subsec:ara_accessibility}
ARA is accessible to the research community through the ARA Portal (\href{https://portal.arawireless.org/}{\texttt{portal.arawireless.org}}) free of cost. Researchers (individuals, research groups, or industries such as AgTech companies) who wish to use the \ara\ testbed are required to abide by the ARA Use Policy~\cite{ARAAccep87:online} and submit a registration request, which will be reviewed by the ARA team before access is granted. To help researchers become familiar with the \ara\ experiment workflow and various wireless platforms, \ara\ provides a webpage~\cite{ARAWirel78:online} describing the testbed features as well as a comprehensive user manual~\cite{ARAUserM67:online}. Once access is granted, users can log in to the \ara\ Portal, reserve resources, and perform the experiments entirely remotely. For AgTech applications and innovation, \ara\ facilitates integration at the device stratum, where agricultural vehicles equipped with sensors and actuators can operate remotely or transmit field data directly to the edge servers at the edge stratum or compute nodes at the cloud stratum depending on the nature of application. This capability is exemplified in Section \ref{subsec:appPilot}, which presents a pilot study demonstrating high-impact applications within \ara.


\section{Implementation of \ara}
\label{sec:implementation}

Based on the key design principles discussed in Section~\ref{sec:design}, we now elaborate on the implementation details of \ara\ BS sites and UE stations, as well as our \araoptical\ platform and approaches to enabling programability of COTS platforms and precision time synchronization.

\subsection{Base Station (BS) Site 
} \label{subsec:BS-site}

\begin{figure}[!tbp]
    \centering
    \includegraphics[width=1.05\columnwidth]{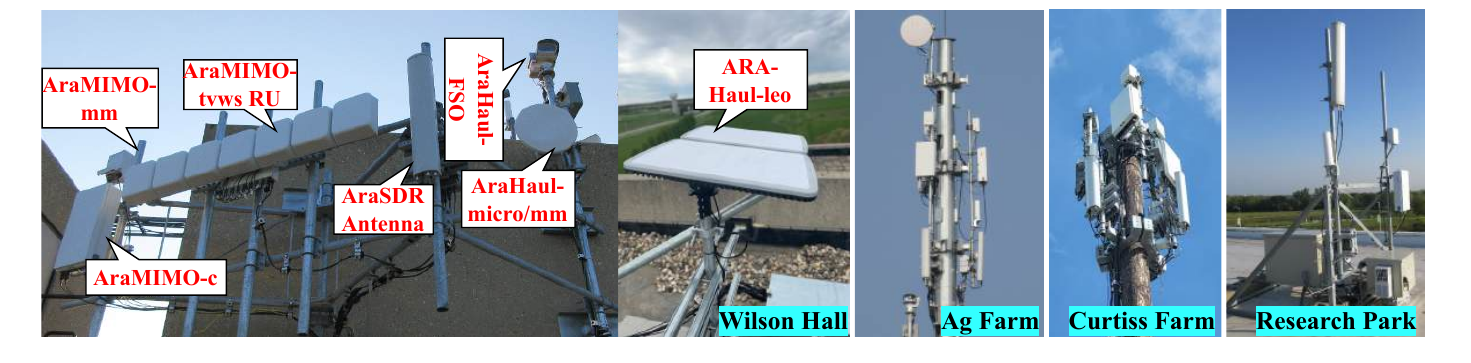}
    \caption{Physical BS installation 
    }
    \label{fig:bs-deployment}
\end{figure}

The BS site has \araran\ BS and \arahaul\ nodes as shown in Figure \ref{fig:system-architecture}. The physical installation of four representative BS sites is shown in Figure~\ref{fig:bs-deployment}. 
The heights and locations of the BSes 
have been carefully chosen through coverage simulations and field testing.
In addition to radio resources, each BS site is equipped with compute resources for 
experiments, as well as resource and experiment orchestration.
For instance, the \aramanagement\ computer for running management and control-related software daemons, e.g., spectrum monitoring, an \arahost\ computer for running \arahaul\ experiments, and \arasdr\ host computer for running the UHD drivers for NI N320 radios at the BS.
Two of the BS sites (i.e., \researchpark\ and \agronomyfarm) are equipped with Edge servers to process data at the edge for real-time 
applications. 

\begin{figure}[!htbp]
    \centering
    \includegraphics[width=0.7\textwidth]{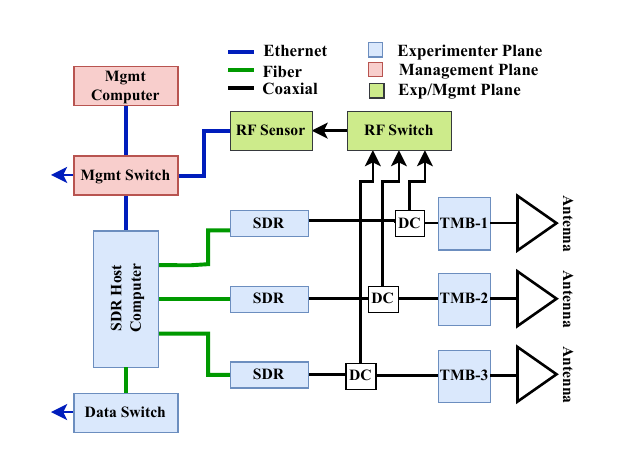}
    \caption{Architecture of field-deployed \arasdr\ BS
    }
    \label{fig:bs-rf}
\end{figure}

\subHeading{\araran\ BS.}
Each \araran\ BS hosts three 
\arasdr~(NI N320~\cite{NI-USRP-N320}). Three \arac~(Ericsson AIR 6419~\cite{Ericsson-Radio}) and three \aramm~(Ericsson AIR 5322) radios are deployed at \wilsonhall, \agronomyfarm, \researchpark\ and \curtissfarm\ with \wilsonhall, \boone, \gilbert\ and \isics\  hosting three \aratvws~(Skylark Faros V2 \cite{Skylark-Faros}) Radio Units~(RUs), enabling a diverse range of operations and experimentation with respect to coverage, capacity, 
and programmability.    
Figure \ref{fig:bs-rf} shows our design of the \arasdr\ BS architecture \cite{arasdr}. 
Each SDR is connected to a band 78 Time Division Duplex~(TDD) Tower Mounted Booster~(TMB) comprising of a Power Amplifier in the transmit chain and Low Noise Amplifier~(LNA) in the receive chain. 
Each TMB is connected to a CommScope SS-65M-R2 antenna. The three CommScope antennas face $120^\circ$ apart.
The Directional Coupler~(DC) and RF~Sensor are for wireless guard as discussed in Section~\ref{subsec:arasoft-design}. 
The SDRs, along with the TDD frontends, support open-source 5G and O-RAN experiments using software platforms such as OpenAirInterface5G~\cite{OpenAirI2:online}. 

\aratvws\ 
BS is a programmable, TVWS many-antenna platform based on the Faros v2 platform of Skylark Wireless \cite{Skylark-Faros}. It follows the CU, DU, RU architecture with one CU, one DU, and three RUs at each BS, 
operating at Low-UHF frequencies from 470--776\,MHz with a target band from 539--593\,MHz. The target band was chosen after field trials to check spectrum occupancy in the coverage area of \ara. 
Each RU is facing $120^\circ$~apart and has a linear array of 14 cross-polarized antennas. 
 
\arac\ BS is based on Ericsson AIR 6419 operating in the n77 5G band (3.45--3.55\,GHz) with 192~antennas per sector, supporting up to 16~layers of MU-MIMO while \aramm\ is based on Ericsson AIR 5322 operating in the n261 frequency band (27.5--27.9\,GHz) with 384~antennas per sector, supporting up to 8 layers of MU-MIMO.
\arac\ and \aramm\ operate in the 5G Stand-Alone~(SA) mode and support New Radio Dual Connectivity (NR-DC) with the mmWave carrier anchored at mid-band carrier, providing an aggregate communication capacity up to 3+\,Gbps enabling advanced rural wireless applications such as smart agriculture and Ag phenotyping.

\subHeading{\arahaul\ Node.} 
\arahaul\ \cite{arahaul} enables long-distance, high-capacity, and affordable x-haul connectivity to rural areas. 
\arahaul\ nodes are installed at \agronomyfarm, \wilsonhall, 
\boone, \gilbert, and \isics, thus forming a wireless x-haul mesh network in real-world rural environment. 
    The exact locations and heights of the \arahaul\ radios are carefully planned to ensure 
    line of sight~(LOS) path. In our study, we conducted thorough LOS testing over distances up to 15\,km using various optical sighting tools, including riflescope, spotting scope, astronomical telescope, and visible laser source. The \arahaul\ platforms are programmable 
    through wrapper APIs as discussed in Section~\ref{subsec:COTS-programmability}. In addition to wireless x-haul nodes, we also deployed \mbox{\araoptical\ }nodes between \wilsonhall\ and \agronomyfarm\ as discussed in Section~\ref{subsec:araoptical} and LEO satellite terminal at \wilsonhall\  from Hughes using the Oneweb satellite network. The \arantn\ link can be accessed from \arahost\ compute node at the BS, enabling measurability of the NTN links and configuring end-to-end paths to carry data from the field application at \agronomyfarm, e.g., drone cameras in the crop farms via \araran\ BS and the \aviatmicro\ link for processing at the edge at \wilsonhall\ and sending the processed data through the \arantn\ link to the data center or Internet. 

\subsection{User Equipment (UE) Station
}
\label{subsec:UE-station}
The UE stratum consists of \araran\ UE and the associated sensing, control, and/or actuation devices. 
\begin{figure}[!htbp]
    \centering
    \includegraphics[width=1.05\columnwidth]{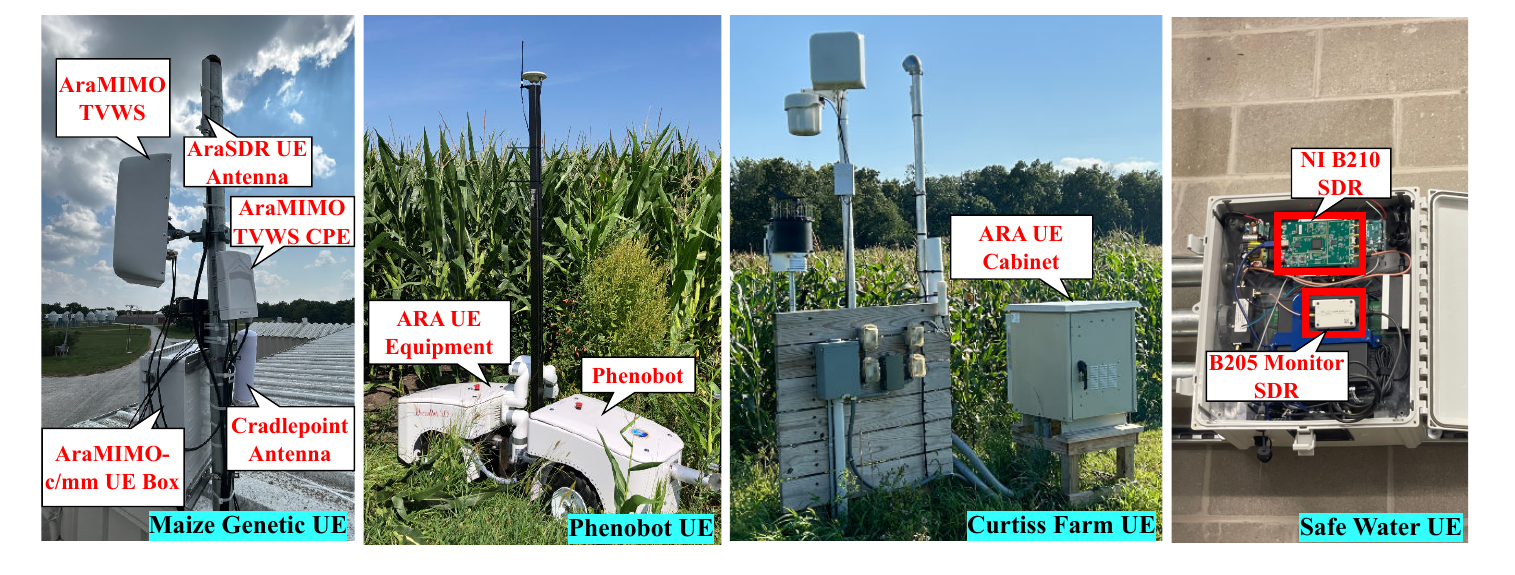}
    \caption{\ara\ field-deployed UEs 
    }
    \label{fig:ue-deployment}
\end{figure}

Each UE has three radios, i.e., USRP B210 SDR, Skylark Wireless Customer Premises Equipment~(CPE), and Quectel COTS 5G UE. 
Each UE has a SuperMicro mini-server 
host computer that runs the 5G protocol stack for NI B210 SDR, software drivers for Quectel COTS UE, and Skylark Wireless application for the Skylark CPE. 
    The host computer is available for experimenters as a compute resource for programmable experiments. 
    \araran\ UE has a Juniper EX2300 switch that ensures connectivity between all elements on the \araran\ UE. 
The management computer runs management and control applications such as spectrum monitoring and remote power management. The USRP B205 SDR is deployed at each UE location for spectrum sensing and spectrum policy enforcement through wireless guard. One Cradlepoint IBR600C cellular router is also installed at each UE station with a commercial LTE SIM card, serving as a backup management channel. 
Figure~\ref{fig:ue-deployment} shows a field-deployed UE with Skylark CPE and antenna, SDR antenna, Cradlepoint router antenna, and COTS UE box deployed on a pole. The figure also shows a mobile UE in a phenotyping robot (phenobot) and a fixed UE deployed in a crop farm. A UE box 
with computers, switch, and SDRs inside is also shown.

\subsection{\araoptical\ Communications System} 
\label{subsec:araoptical}

\begin{figure}[h]
    \centering
    \includegraphics[width=0.9\columnwidth]{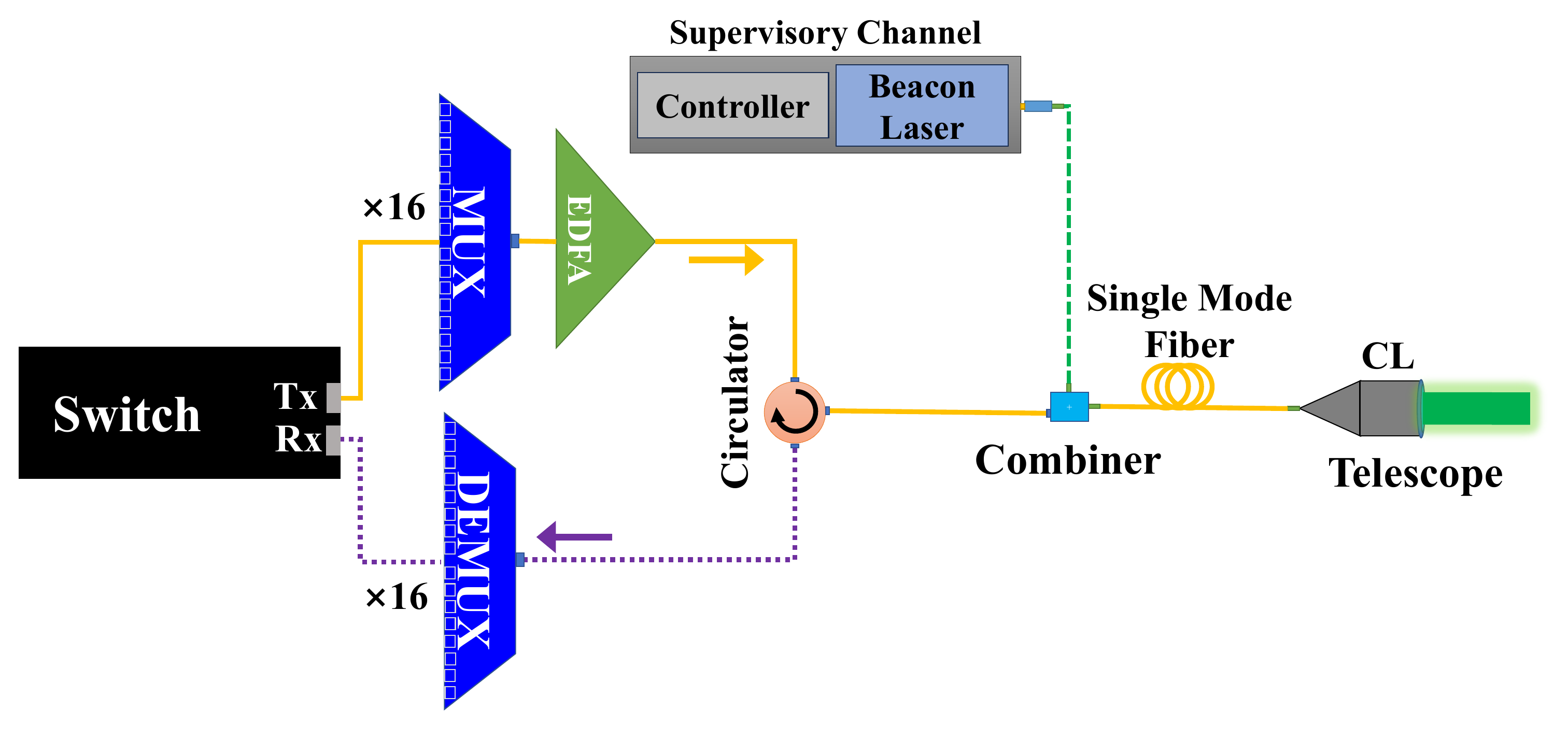}
    \caption{Architecture of \araoptical\  signal branch}
    \label{fig:araOptical_architecture}
\end{figure}

\araoptical\ is a first-of-its-kind long-distance, high-capacity terrestrial free-space-optical-communication (FSOC) platform, capable of achieving up to 160\,Gbps data rate over 10\,km aerial distance in rural environments. Figure~\ref{fig:araOptical_architecture} 
shows the architecture of the \araoptical\ signal branch. Each node employs 16 wavelength-division-multiplexing (WDM) channels, each using an SFP+ transceiver operating at 10\,Gbps. These transceivers connect to an optical module with an erbium-doped-fiber-amplifier (EDFA) for signal amplification. The amplified signal beam is transmitted over-the-air using an optical collimator~(CL) with a 35\,$\mu$rad divergence angle, significantly reducing geometric losses. The Circulator separates the received signal from the transmitted signal 
and forwards the received signal to the WDM device for demultiplexing. 
    To enable LOS FSOC at long distance, the \araoptical\ platform has a carefully designed system for self-alignment using mechanisms such as the beacon signal which has exactly the same LOS path as the data signal but has much wider beamwidth and thus easy to align.

\araoptical\ offers first-of-its-kind terrestrial FSOC measurability and programmability through wrapper APIs. 
These APIs enable the measurement of parameters such as 
reception power, throughput, and communication latency from switch interfaces. The APIs also enable controls such as 
transmission power and bitrate adaptation, beam steering, and routing. 


\subsection{Programmability for COTS Platforms 
} 
\label{subsec:COTS-programmability}

To enable programmability of COTS platforms while leveraging their performance and robustness in wireless and application experiments, we develop wrapper APIs that expose system configuration and control according to experiment requirements while ensuring the system integrity from potential experiment misbehavior. 
    More specifically, for the \aratvws\ platform~\cite{aramimo} with hardware and software access control, we define modding 
APIs capable of executing custom user scheduling and grouping algorithms, along with wrapper APIs for the configuration and control of the TVWS massive MIMO BSes and UEs. For the \arac, \aramm, and \arahaul\ platforms, we define wrapper APIs around the device-level functions to enable programmability and measurability while ensuring safety through strict access control. 
    Besides \araran\ and \arahaul, \ara\ includes resources (e.g., weather sensors and PDUs) which do not need explicit resource reservation for their use in experiments, that is, the devices can be polled for weather and power-related measurements concurrently by multiple users at any given time. A dedicated server along with custom APIs are developed for users to collect the most recent as well as historical weather and power consumption information. 
The wrapper APIs for each platform are containerized and made available to \ara\ users for research experiments. 

\subsection{Precision Time Synchronization}

Time synchronization plays a crucial role in the operation and management of networks, particularly in the context of Ultra-Reliable, Low-Latency Communications (URLLC), essential for next-generation wireless systems such as 5G, 6G, and Open RAN. Within the \ara\ platform, we have developed AraSync \cite{arasync}, an end-to-end solution dedicated to facilitate advanced wireless experiments and applications that demand precise time synchronization. Leveraging Precision Time Protocol (PTP), AraSync achieves synchronization accuracy down to the nanosecond range. Integrated with fiber networks, AraSync extends its capabilities to synchronize timing across the AraHaul wireless x-haul network, which encompasses long-range, high-capacity mmWave and microwave links.



\section{\ara\ Capacity, Coverage and Example Experiments} 
\label{sec:EvalExperiments}

To demonstrate wireless research and innovation uniquely enabled by ARA (see Section~\ref{sec:ara-enabled-research}), here we characterize the \ara\ capacity and coverage, and we showcase examples of ARA-enabled experiments in advanced wireless and applications.

\subsection{\ara\ Capacity and Coverage 
} \label{subsec:characterization}

\subsubsection{\araran\ Capacity Characterization}

To characterize the capacity of the heterogeneous \araran\ platforms, we measure 
the link capacity in the south sector of the \wilsonhall\  BS. The data collection involves simultaneous measurements at several distances from the BS. The \aramm, \arac\ and \arasdr\ capacity is measured using \emph{iPerf3} while the capacity of \aratvws\ is measured using the software application at BS which calculates the achievable capacity based on the uplink and downlink SNRs. \aramm\ and \arac\ are transmitting at 128\,W while \aratvws\ is transmitting at 10\,W, which is the maximum power supported for each platform. \arasdr\ is transmitting at 10\,mW which is the optimal transmit power for maintaining stable open-source 5G connections between SDR-based UEs and BSes in \ara\ outdoor environments using the development branch of OpenAirInterface5G. 
The measurements are taken starting at 170\,m up to 8.6\,km from the BS. The first six points are roughly 100\,m apart while the remaining points are approximately 350\,m apart. Several crop and livestock research farms are located along this route,
making it an ideal route for rural- and ag-focused study.

\begin{figure}[!htbp]
    \centering
    \hspace{-5mm}\includegraphics[width=.8\textwidth]{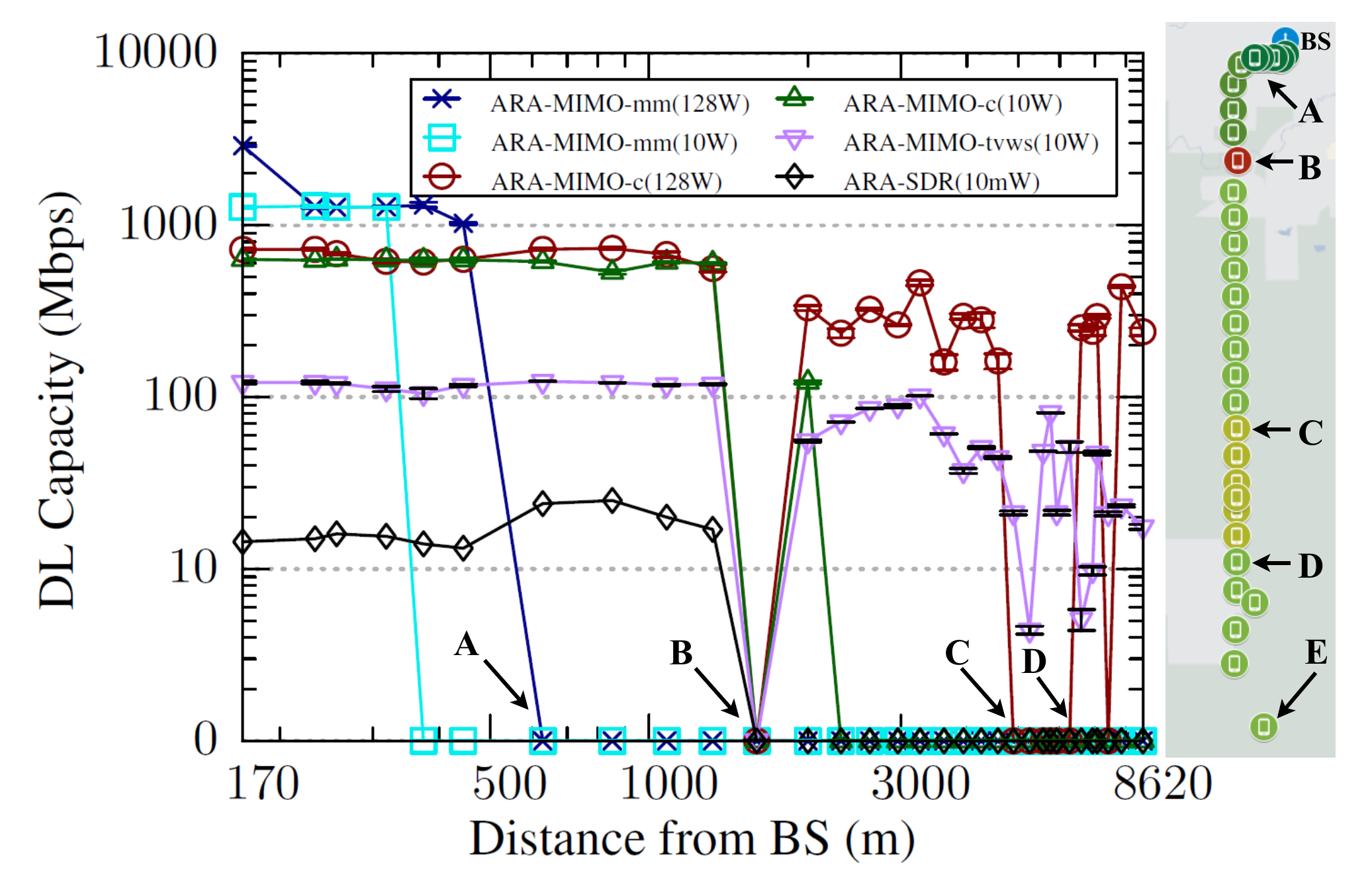}
    \caption{Comparison of coverage and capacity of \aramm, \arac, \aratvws, and \arasdr\ platforms.
    }
    \label{fig:ran_comparison}
\vspace{-5mm}
\end{figure}

Figure~\ref{fig:ran_comparison} shows that \aramm\ provides the maximum downlink~(DL) capacity of 3\,Gbps with bandwidth of 400\,MHz 
and coverage up to 500\,m (Point A) from the BS. \arac\ and \aratvws\ achieve the maximum DL capacity of 650+\,Mbps with 100 MHz bandwidth and 120\,Mbps with 24\,MHz bandwidth respectively, having decent coverage up to 8.6+\,km. 
The use of \arasdr\ is primarily for its full programmability and not for capacity, but \arasdr\ still enables a capacity of 25+\,Mbps with 40\,MHz bandwidth and up to 1.2+\,km distance. 
About 1.6\,km from the BS (Point~B), the terrain has a severe dip in elevation which prevents UEs from  connecting to the BS due to severe signal blockage.  There exist some locations between Points C and D~(about 4.5--6\,km from the BS) where \arac\ shows no coverage due to the blockage from trees and farm buildings, however, \aratvws\ can still connect to the BS. 
As the elevation goes up 
around 6.5\,km (Point~D),
the \arac\ UE is able to reconnect to the BS.
We see that there is almost uniform coverage up to 3\,km distance and then there exists a transition region where the capacity varies due to the terrain conditions. For characterizing the heterogeneity of \ara, 
we also evaluate the \arac\ and \aramm\ BSes with the same transmission power as \aratvws~(i.e., 10\,W). We see 
that the coverage is reduced to 2\,km and 300\,m for \arac\ and \aramm, respectively. 

\subsubsection{\araran\ Coverage Characterization}

For understanding opportunities for wireless and application co-evolution studies, we also characterize the wide coverage  of the COTS platforms \arac\ and \aratvws.  
For \arac, we perform drive tests along the roads in the coverage area of the four \arac\ BSes using a Nemo Handy UE ~\cite{NemoHandy:online}. For \aratvws, we use \wilsonhall\ BS with UE at different locations, roughly 1.6~km apart. We use the maximum transmit power for \arac\ and \aratvws\ BSes (i.e., 128\,W and 10\,W, respectively). 
Measurement data from the field is analyzed using Laplace-equation-based, Physics-Informed Neural Network~(PINN) \cite{PINN} model. The model consists of a neural network with five fully connected layers, as we designed in~\cite{rural-spectrum}, which 
takes \textit{only the geographical coordinates}  of sample points from the dataset to learn the coverage variability over space. The predicted value from the output layer of the neural network is compared with the measured throughput to calculate the data error. In PINN, along with the data error, loss functions derived from Partial Differential Equations~(PDEs) representing the Laplace equation of the underlying physics are incorporated to adjust the neural network's weights during training. The inclusion of PDE-driven loss significantly improves the prediction accuracy, as we observed in~\cite{rural-spectrum}. 

\begin{figure}[h]
  \centering
\includegraphics[width=\columnwidth]{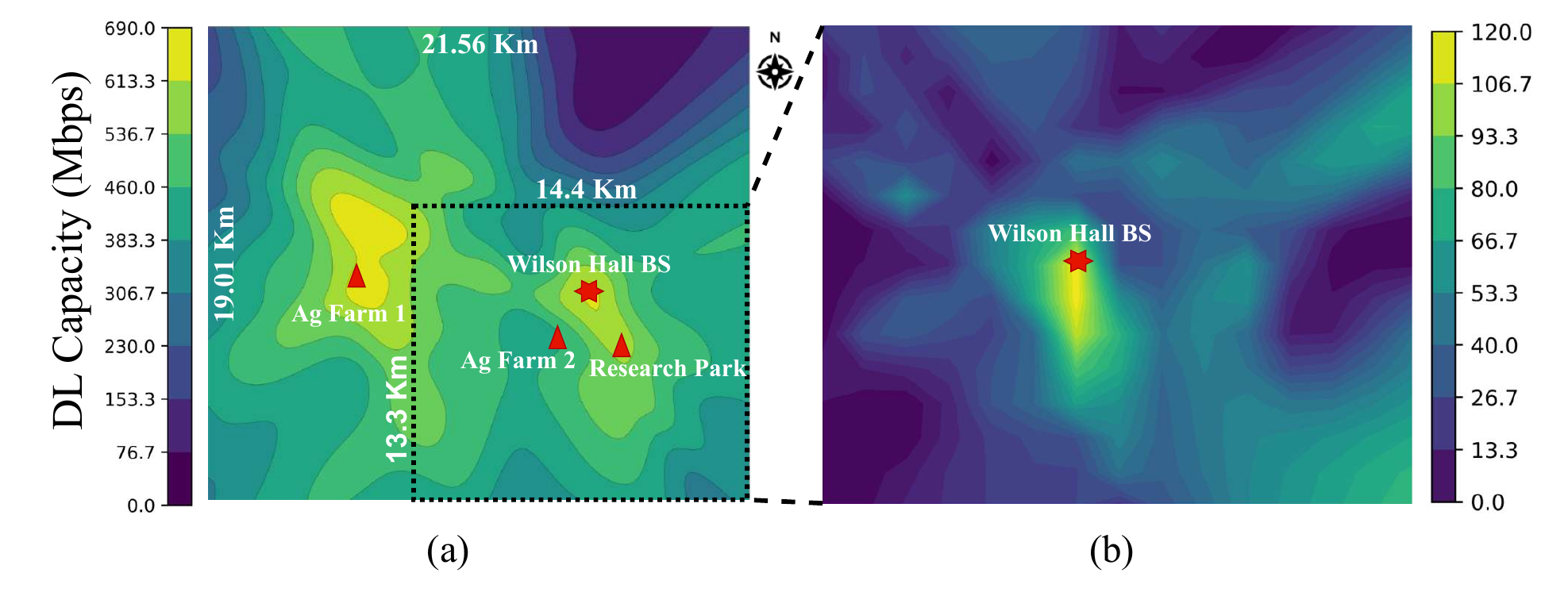}
  \caption{\arac~(a) and \aratvws~(b) Coverage map.
  }
  \label{fig:ara-coverage}
\end{figure}

The coverage map generated with PINN for \arac\ is shown in Figure~\ref{fig:ara-coverage}a, where the yellow and green points indicate higher and medium capacity, respectively, while the dark blue points represent limited-to-no coverage. 
The coverage grid is 19\,km $\times$ 21.5\,km  and shows good coverage near the BSes. We observe varying capacity as we go farther from the BSes and zero coverage toward north-east due to the blockage from terrain and residential areas. 
For \aratvws, 
the coverage map is shown in Figure~\ref{fig:ara-coverage}b where the size of the grid is 11.7\,km $\times$ 14.4\,km, a sub-area of Figure~\ref{fig:ara-coverage}a. 
The figure shows good coverage toward the south covering the rural crop and livestock farms, and the grain mills. Some spots on the grid toward north and north-west are not covered due to the blockage from residential buildings in the city, and the locations towards far-west have blind spots due to the terrain and the angle of UE from the BS sector-antennas. A few locations toward east are also not covered due to severe blockage. 

\subsubsection{\arahaul\ Capacity Characterization}

\arahaul\ features heterogeneous, long-distance wireless x-haul platforms across the 
microwave, 
mmWave, 
and free-space-optical 
frequency bands, enabling the use of spectral diversity for performance and robustness. 
We use the 10.15\,km long \arahaul\ link between \agronomyfarm\ and \wilsonhall\ to characterize the \arahaul\ capacity and heterogeneity. 

\paragraph{\aviatmicro\ and \aviatmm\ Link Capacity}

Out of the three bands, the most robust one is the microwave band at 11\,GHz.
We 
observe a link throughput as high as 892\,Mbps (i.e., 97.2\% of the theoretical limit of 918\,Mbps) thanks to the robust nature of wireless communications at the microwave band. 
In comparison, the mmWave band at 80\,GHz is more sensitive and susceptible to weather. 
While the theoretical limit of the mmWave link throughput is 10\,Gbps,
the highest throughput observed in our experiments is 6.36\,Gbps, as shown in Table~\ref{tab:Arahaul_performance}.

\begin{table}[!htbp]
\caption{\aviatmicro\ and \aviatmm\ link capacity
}
\centering
\label{tab:Arahaul_performance}
\resizebox{\columnwidth}{!}{
\begin{tabular}{|l|l|l|l|l|l|l|l|l|}
\hline
Link                      & \begin{tabular}[c]{@{}l@{}}Freq.\\ (GHz)\end{tabular} & \begin{tabular}[c]{@{}l@{}}BW\\(MHz)\end{tabular}     & \begin{tabular}[c]{@{}l@{}}Mod.\\(QAM)\end{tabular} & \begin{tabular}[c]{@{}l@{}}Tx-Power\\(dBm)\end{tabular} & \begin{tabular}[c]{@{}l@{}} Antenna\\Gain~(dBi)\end{tabular} & \begin{tabular}[c]{@{}l@{}}Beam-\\ Width\end{tabular} & \begin{tabular}[c]{@{}l@{}}Achieved\\ Capacity\end{tabular} & \begin{tabular}[c]{@{}l@{}}Theoretical\\ Capacity\end{tabular}  \\ \hline\hline
\aviatmicro & 11                                          & 100  & 4096   & 26                                              & 33.6                                                        & 3.2\degree                             & 892~Mbps                             & 918~Mbps                                               \\ \hline
\aviatmm    & 80                                          & 2000 & 32     & 13                                              & 50                                                          & 0.5\degree                             & 6.36~Gbps                             & 6.82~Gbps                                              \\ \hline
\end{tabular}}
\end{table}

\paragraph{\araoptical~Link Capacity}

\araoptical\ is a terrestrial long-distance FSOC link, a unique feature available in \ara. 
We observe in our field measurement that \araoptical\ transceivers can attain a peak reception power of -6.86\,dBm, which is about 17\,dB above the receiver sensitivity of -24\,dBm, allowing for a maximum communication capacity of 10\,Gbps per signal branch and 160\,Gbps aggregate capacity with 16 signal branches. 
Our \araoptical\ APIs 
allow researchers to measure, for the first time, the scintillation effect and weather impact on such terrestrial long-distance FSOC links in real-world settings, and they enable experimentation with FSOC systems and protocol designs. 
Refer to Section~\ref{subsec:xhaul_experiment} for example studies.

\subsection{Example Rural Wireless Experiments} \label{subsec:wirelessExpt}

In what follows, we demonstrate unique exemplar wireless experiments enabled by \ara. 

\subsubsection{Weather Experiments}

\begin{figure}[!htbp]
    \centering
    \includegraphics{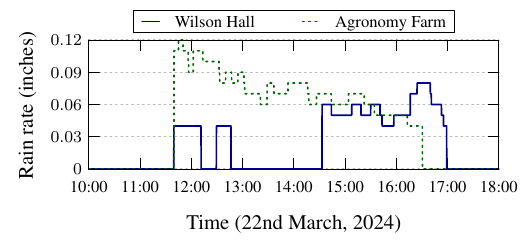}
    \caption{Example temporal and spatial variation of rain rate collected in \ara}
    \label{fig:weather_data}
\end{figure}

\ara\ is unique in its capability of enabling advanced experiments to study the impact of weather on wireless communication systems. To provision the weather data, \ara\ is equipped with high-precision weather sensors and disdrometers to capture fine-grained variations in weather conditions such as precipitation type (rainy, snowy, foggy, or drizzling), rain drop size and velocity distribution, wind speed and direction, humidity, temperature, and atmospheric pressure. 
With the weather data, \ara\ offers a new dimension to the wireless research to unravel the possible hidden correlation between the weather conditions and the wireless system performance. Such a study is critical in analyzing the performance of \araran\ and \arahaul\ links, especially the free space optical link, which has significant dependence on atmospheric conditions. Besides predicting the weather, the radio parameters can be tuned based on the anticipated weather conditions to achieve optimum performance. Figure~\ref{fig:weather_data} 
shows an example of the temporal and spatial variation of rain rate collected from two BS sites. The spatial diversity of weather conditions indicates the possibility of location-based tuning of the device attributes considering the weather condition at the location.  We demonstrate the influence of rain on the \araran\ and \arahaul\ performance in Section~\ref{subsec:packet_delay_analysis}
and Section~\ref{subsec:xhaul_experiment}, respectively. A more detailed study on rural wireless channel measurements can be found in \cite{ara-rural-channel-study}. 

\subsubsection{Communication Delay Analytics Using SDRs and Open-Source 5G Platforms}
\label{subsec:packet_delay_analysis}

\ara\ uses fully programmable USRP SDRs N320 and B210 at the BSes and UEs, respectively. Together with open-source 5G platforms such as OAI, these SDRs enable deep insight into the system behavior. For instance, one open challenge in 5G and 6G is how to enable real-time cyber-physical systems that 
require data packets to be delivered by their deadlines. One important step toward meeting such requirements is to understand the sources of delays. 
    To this end, \ara's field-deployed, fully-programmable SDRs enable collection of fine-grained delay data for packets traversing the 5G network stack using the OpenAirInterface5G software platform. 

More specifically, using 
OAI 5G software stack, we extend \textit{LatSeq}~\cite{9417345} (a low-impact internal delay measurement tool for OAI 5G) to capture, store, and analyze the internal delays of packets between UEs and gNBs for both uplinks and downlinks. 
We insert measurement points across the whole OAI 5G network stack from the application layer to the physical layer. The measurement data include timestamps of packets 
entering and exiting each layer. 
These 
data are crucial in rebuilding the packet journey and estimating the per-layer contribution to the end-to-end delay of a packet.

For an outdoor SDR link whose length is 200\,m, we leverage the \texttt{ping} utility to generate packets in both the uplink and downlink directions and collect the packet delay data in the form of \texttt{.lseq} logs from the gNB and UE. 
    Given the significant impact of weather on wireless communications, 
    we investigate the impact of weather on packet delays. 
In particular, we measure packet communication delay with and without rain. We use a channel bandwidth of 40\,MHz, 6 downlink and 4 uplink OFDM symbols, 30\,kHz subcarrier spacing, and 3586.40\,MHz center frequency
in this experiment.

\begin{figure}[ht!]
  \centering
\includegraphics[width=0.7\textwidth]{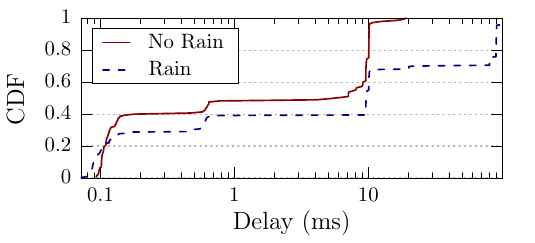}
  \caption{CDFs of packet delays in different weather
  }
  \label{fig:raincdf}
\end{figure}

Figure~\ref{fig:raincdf} shows the cumulative distribution function (CDF) of the delays of 100 packets generated using \texttt{ping}. Each packet is 100\,KB in size and generated every 10\,ms. It can be seen that, given a target delay bound of 10\,ms,
almost all packets meet this requirement in the absence of rain. However, with rain falling at an average rate of 2.06\,inches per hour,
only 70\% of packets meet the delay bound. 

To understand the above behavior, we show in Table~\ref{tab:layercon} the delays introduced by the individual layers of the 5G network stack. 
\begin{table}[t!]
\centering
\footnotesize
\caption{Mean layer contributions of total packet latencies}
\begin{tabular}{l|cl|}
\cline{2-3}
\multirow{2}{*}{}                   & \multicolumn{2}{c|}{\textbf{Average delay\,(msec) $\pm$CI(95\%)}}                      \\ \cline{2-3} 
                                    & \multicolumn{1}{c|}{\textbf{Rain}}        & \multicolumn{1}{c|}{\textbf{No Rain}}     \\ \cline{2-3}
                                    \hline
\multicolumn{1}{|l|}{\textbf{SDAP}} & \multicolumn{1}{l|}{0.00391 $\pm$0.00115} & 0.00355 $\pm$0.00029                      \\ \hline
\multicolumn{1}{|l|}{\textbf{PDCP}} & \multicolumn{1}{l|}{0.00690 $\pm$0.00017}  & 0.00712 $\pm$0.00016                      \\ \hline
\multicolumn{1}{|l|}{\textbf{RLC}}  & \multicolumn{1}{l|}{30.0579 $\pm$0.28633} & 7.63437 $\pm$0.11103                      \\ \hline
\multicolumn{1}{|c|}{\textbf{MAC}}  & \multicolumn{1}{c|}{0.16896 $\pm$0.00010}  & \multicolumn{1}{c|}{0.08133 $\pm$0.00014} \\ \hline
\multicolumn{1}{|c|}{\textbf{PHY}}  & \multicolumn{1}{c|}{0.01896 $\pm$0.00108} & \multicolumn{1}{c|}{0.01784 $\pm$0.00111} \\ \hline
\end{tabular}
\label{tab:layercon}
\end{table}
We see that the average delay introduced by the MAC layer doubled in the presence of rain. This is due to an increase in the number of packet re-transmissions during rain. 
    Another layer contributing to the large  delay during rain is the Radio Link Control~(RLC) layer. This is due to the use of lower-order modulation-and-coding-schemes during rain. 
Accordingly, much smaller Transport Block Sizes~(TBS) are used, which in turn leads to more time spent for segmenting packets at the RLC layer and,  more MAC layer frames have to be transmitted, resulting in longer delay. 
Such measurement insight helps suggest optimization strategies toward reducing communication delays, for instance, by 
increasing the number of scheduled physical-resource-blocks 
to increase the TBS at the RLC layer which in turn reduces segmentation. 
 
Similarly, with the SDRs deployed in real-world outdoor rural settings in addition to a 50-node indoor sandbox, ARA serves as a first public O-RAN testbed for experiments tailored to rural and agriculture use cases \cite{ara-o-ran}.

\subsubsection{User Grouping Experiment in TVWS mMIMO 
}

\begin{figure}[!htbp]
    \centering
    \begin{subfigure}[b]{0.8\textwidth}
        \centering
        \includegraphics[height=2.1in]{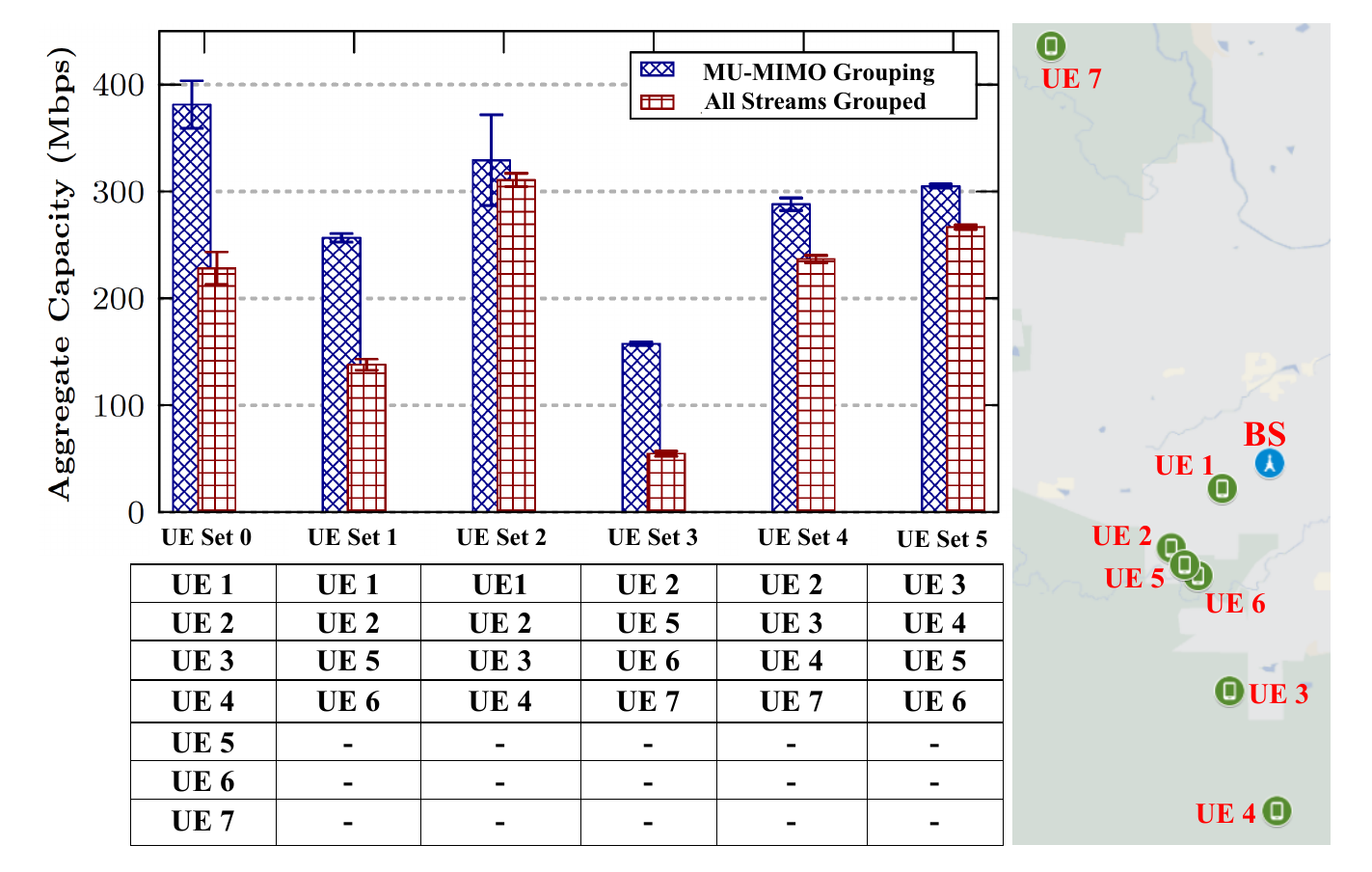}
        \caption{Effect of user grouping on aggregate capacity}
        \label{fig:grouping_capacity}
    \end{subfigure}
    
    \vspace{0.2cm} 

    \begin{subfigure}[b]{0.8\textwidth}
        \centering
        \includegraphics[scale=1]{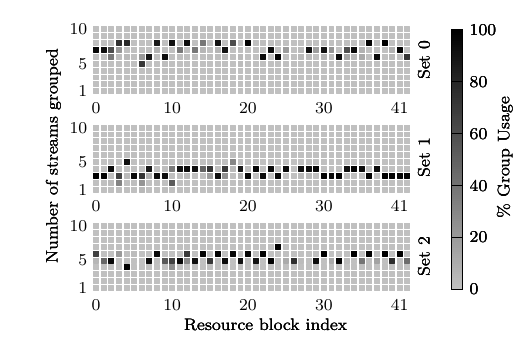}
        \caption{Groups scheduled with $n$ streams}
        \label{fig:grouping_streams}
    \end{subfigure}
    
    \vspace{0.2cm} 

    \begin{subfigure}[b]{0.8\textwidth}
        \centering
        \includegraphics[scale=0.8]{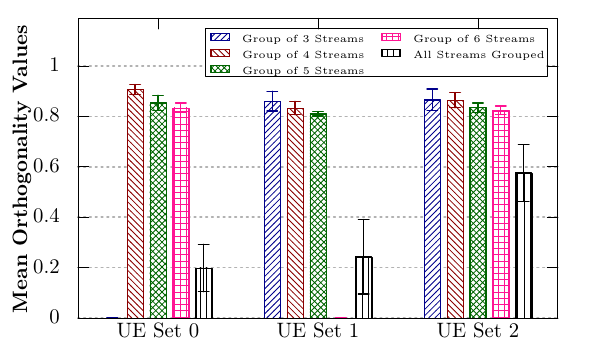}
        \caption{Mean orthogonality values of different groups}
        \label{fig:orthog_values}
    \end{subfigure}
    
    \caption{Effect of spatial distribution of users on MU-MIMO capacity}
    \label{fig:stacked_figures}
\end{figure}

Through its rich APIs for massive  MIMO (mMIMO) control and measurement, \aratvws\ 
enables unique experimental studies in the TVWS band at the PHY, MAC, and higher layers in rural settings. 

As an example, we experimentally characterize 
the effect of the spatial distribution of UEs on the achievable capacity. Figure~\ref{fig:grouping_capacity} shows the aggregate capacity achieved when different sets of UEs are connected to the BS. Set~0 includes all seven UEs, while Set~1 to Set~5 have four different UEs each, as can be seen on the map and table in Figure~\ref{fig:grouping_capacity}. Each UE has two RF chains connected to a cross-polarized directional antenna, hence supporting two spatial data streams. Multi-user MIMO~(MU-MIMO) is employed by the BS to group multiple streams together.
    Maximum aggregate capacity of up to 400\,Mbps is achieved with Set 0, where all seven UEs (up to 14 spatial data streams) are connected to the BS, and multiple sub-groups of streams are created and scheduled at different resource blocks~(RBs). 
There is a total of 42 RBs, 540\,kHz each, equivalent to 24\,MHz bandwidth with guard bands. 

Figure~\ref{fig:grouping_streams} shows how groups with different number of streams~($n$) are scheduled at each RB, where dark-gray scales represent more frequent usage of the corresponding group.
The group usage for Set 0 shows that the BS is able to group up to eight streams (out of the available~14) at each RB. For comparison purposes, we also include in Figure~\ref{fig:grouping_capacity} the aggregate capacity achieved when forcing all streams to be grouped together---a non-optimal grouping strategy hence  resulting in a lower capacity. 
Set~1 consists of UEs 1, 2, 5, and~6, as shown on the map in Figure~\ref{fig:grouping_capacity}. This set of UEs achieve an aggregate capacity of 250\,Mbps, which is lower than that achieved by the Set~2 UEs~(340\,Mbps). The reason is that Set~2 UEs are more spatially separated 
 while Set~1 UEs are packed in a smaller area. 
This effect is further illustrated in Figure \ref{fig:grouping_streams} where the system can schedule groups with up to eight streams for Set~0, six streams for Set~2, and up to four streams for Set~1 on most of the RBs.

In Figure~\ref{fig:orthog_values}, we plot the orthogonality values between the channels associated with the grouped streams, which is calculated as one minus the multiple correlation coefficient between them. The figure shows that Set~2 UEs can achieve high orthogonality values with up to six streams, while for Set~1, if six or more streams are grouped together, the orthogonality value is very low and hence not recorded. We also observe that, when all streams are grouped together in each set, the orthogonality value of Set~0 (with 14 streams) is the lowest, while Set~2 (with eight streams) yields a higher orthogonality value than Set~1. This highlights the importance of having a proper grouping strategy for a given set of UEs in order to maximize the overall capacity.

In the above study, the default grouping strategy of the COTS application has been used. The \aratvws\  APIs 
allow experimenters to perform other mMIMO experiments, for instance, those with specific sets of users, specific grouping strategies, 
or specific transmission strategies.
This holds practical implications for rural applications, such as maximizing per group capacity in precision agriculture e.g. with Unmanned Ground Vehicle (UGV) swarms sending sensors data from crop fields. Furthermore, rural applications encompass diverse needs including URLLC, massive Machine Type Communications (mMTC), and enhanced Mobile Broadband (eMBB), thereby enabling, experimental studies on user grouping based on service classes or across different slices sharing the same resource blocks.

\subsubsection{Spectrum Sensing Experiment}

RF spectrum is fundamental to wireless communication systems, acting as their lifeblood. \ara\ has deployed Keysight's RF sensors at the BS sites and NI Ettus SDRs at UE stations to 
capture the spatio-temporal dynamics of RF spectrum within the region.
One interesting spectrum that is worth in-depth investigation is the TV White Space~(TVWS) spectrum from 470 to 698\,MHz. The primary users of this spectrum band are the TV broadcasters and the unlicensed microphones that transmit in this band. It is well-known that the majority of the TVWS is unoccupied or unused for most of the time, which is especially true in rural regions. Therefore, there are abundant opportunities for secondary users to utilize the TVWS spectrum. In order to do so, it is critical to have a thorough understanding of the usage pattern of the primary users, both temporally and spatially. 
We have developed APIs that allow users to collect and study the usage of the TVWS spectrum
near the \ara~footprint.

\begin{figure}[!htbp]
        \centering
        \includegraphics[width=0.6\textwidth]{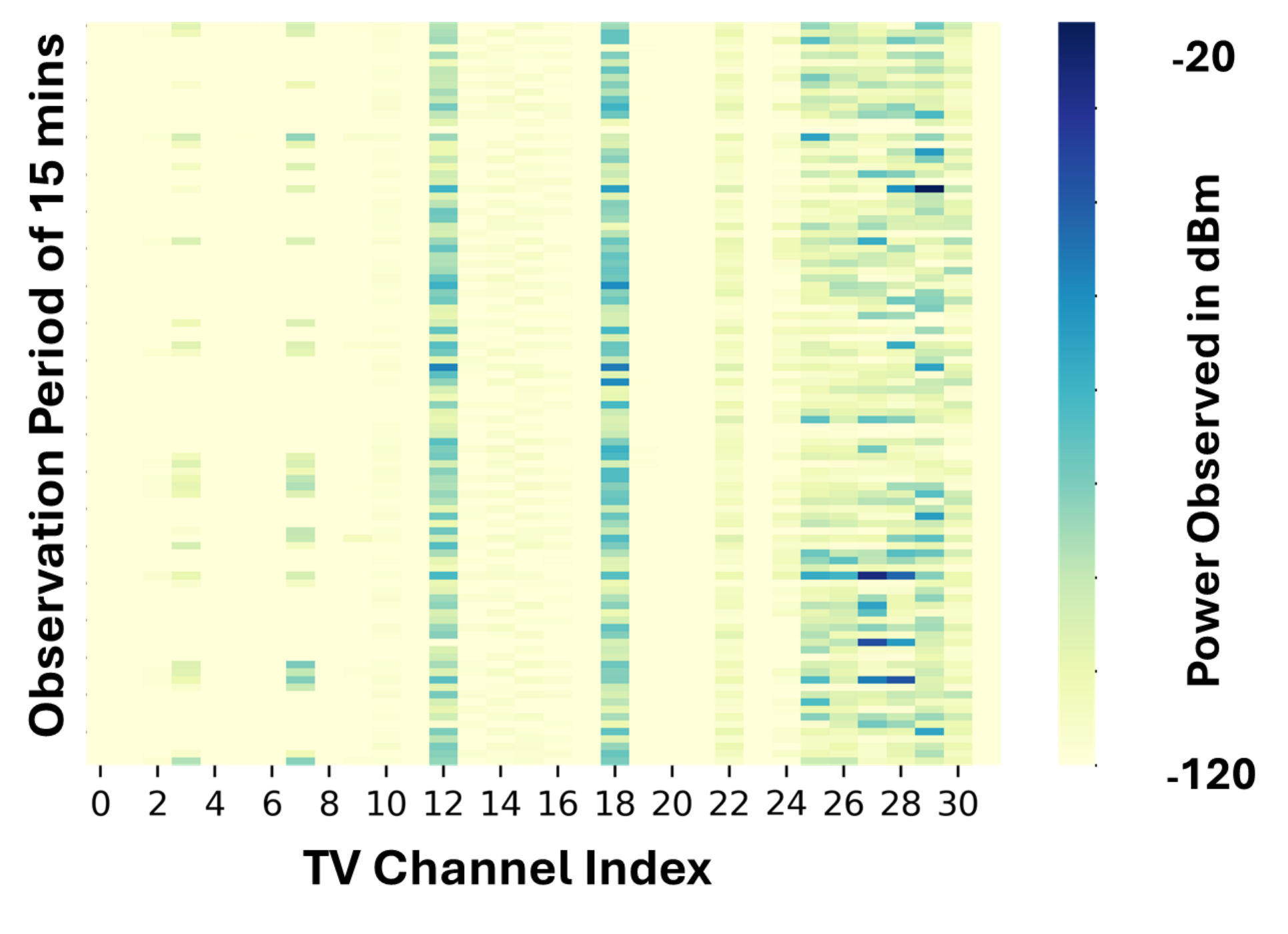}
        \caption{Example TVWS usage near the \ara\ footprint}
        \label{fig:TVWS spectrum}
\end{figure}

An example collection of TVWS usage data is plotted in Figure~\ref{fig:TVWS spectrum},  
where the dynamic behavior of the primary user's transmission activity vividly portrays the spectrum's fluctuating nature. Here, the x-axis denotes the indices of broadcast TV channels where each channel is 6\,MHz wide, and the span ranges from 470\,MHz to 698\,MHz, while the y-axis represents the temporal occupancy of these TV channels for a monitoring time period of 15~minutes where each time slot 
is one second long. In addition, the color intensity of each slot represents the received signal strength captured at the monitoring node ranging from -20 to -120\,dBm.

We have a couple of observations from the captured data. First, as many as 15 TV channels are available near  the \ara~footprint. This available spectrum may be used 
by small-scale Internet Service Providers (ISPs) or Mobile Network Operators (MNOs) to provide a better wireless coverage in rural areas. In addition, it also may be used by farmers to deploy their own IoT networks to help produce better agricultural yield.
Second, the available spectrum in the TVWS is highly fragmented, varying drastically over both time and space, which calls for innovative solutions for better spectrum modeling, management, utilization, and sharing.

With \ara, users can use the provided APIs to run spectrum sensing experiments to study the spectrum of their interest and collect spectrum data at desired time and location.
Captured spectrum data can be used to model the spatio-temporal dynamics of the spectrum occupancy and channel behaviors~\cite{rural-spectrum}. These models serve as valuable resources for the development of spectrum sharing or co-existence protocols and systems, which can then be evaluated on \ara.

\subsubsection{Long-Distance Wireless x-haul Experiment}
\label{subsec:xhaul_experiment}

One key barrier to the real-world adoption of long-distance wireless x-haul communications is the weather impact on communication performance and robustness. Therefore, a critical need in long-distance rural wireless x-haul research is to characterize weather impact, whose insight will shed light on solution avenues. 
    In what follows, we present example experiments characterizing weather impact on heterogeneous wireless x-haul platforms in \ara, in hope of sharing a glimpse into the broad research opportunities enabled by \arahaul.

\paragraph{\aviatmicro~and \aviatmm}

\arahaul\ provides rich APIs for configuring and measuring the \aviatmicro\ and \aviatmm\ links through 
standard NETCONF and SNMP protocols. 
Using these APIs, we measure the received-signal-level~(RSL) and throughput for the following 
link configurations: 
1) \textit{11\,GHz \aviatmicro~link:} 100\,MHz channel bandwidth, 4096\,QAM modulation, and 26\,dBm transmission power;
2) \textit{80\,GHz \aviatmm~link config.~\#1:} 1\,GHz, 16\,QAM, and 14.5\,dBm;
3) \textit{80\,GHz \aviatmm~link config.~\#2:} 2\,GHz, 32\,QAM, and 13\,dBm.

\begin{figure}[!htbp]
    \centering
    \includegraphics[width=0.9\textwidth]{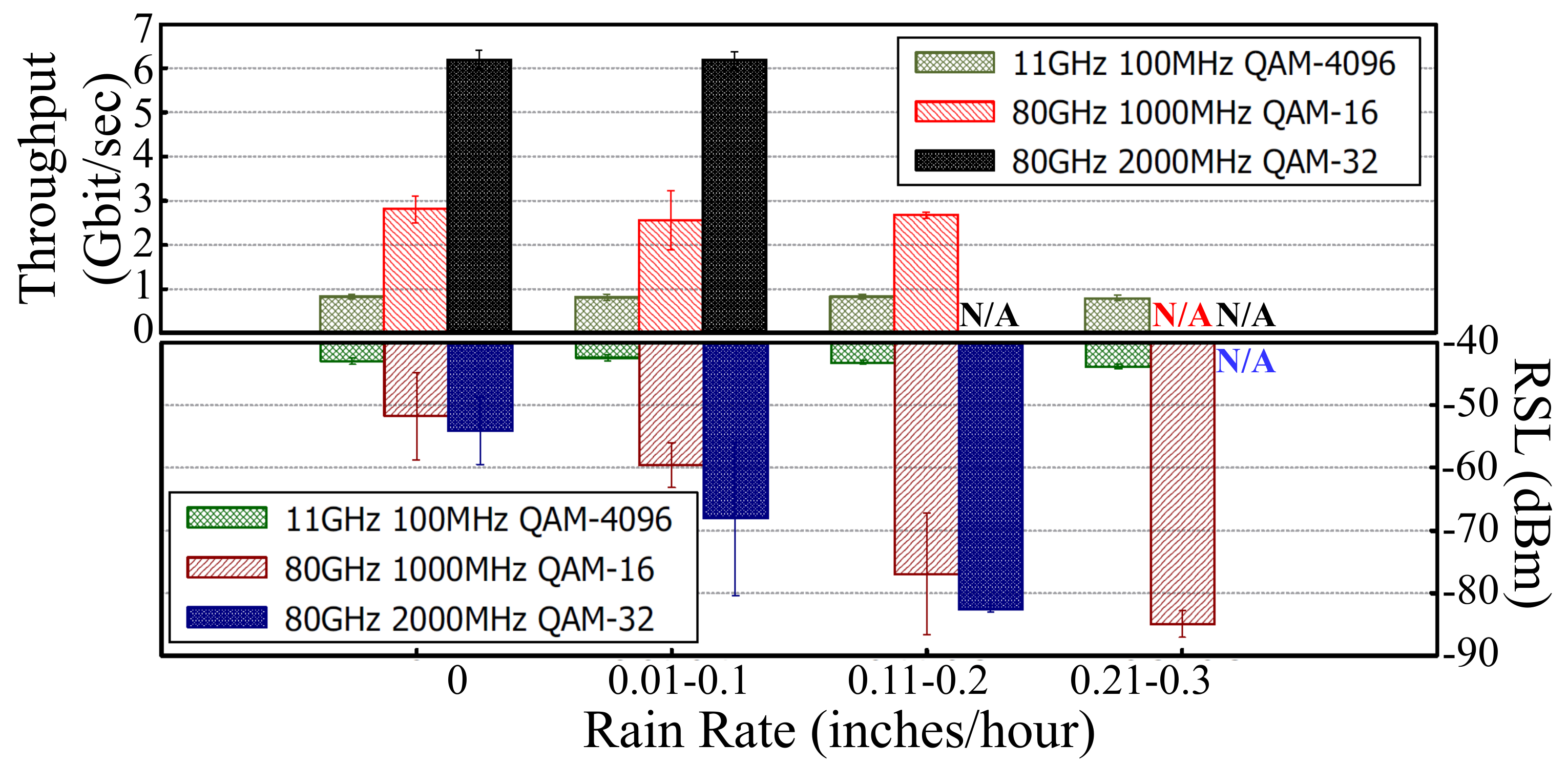}
    \caption{Impact of rain rate on \aviatmicro\ and \aviatmm\ links 
    ($\text{confidence interval}=95\%$) 
    }
    \label{fig:aviat_rain_rate}
\end{figure}
Figure~\ref{fig:aviat_rain_rate} shows the link RSL and throughput 
under different rain rates.
    We see that 
    the microwave link at 11\,GHz is able to maintain steady RSL and throughput regardless of the rain condition. In contrast, the mmWave link at 80\,GHz is more susceptible to weather;  
it yields a significantly higher throughput under no-rain and light-rain conditions, however, its performance deteriorates quickly when the rain rate increases. 
    These results demonstrate the inherent tradeoff between communication capacity and weather resiliency across wireless x-haul links at different frequency bands, as well as the criticality of optimally controlling the operation parameters of higher-frequency 
    x-haul links. They demonstrate the need for leveraging spatiotemporal diversity to enhance overall system robustness, whose study is enabled by the mesh deployment of \arahaul; they also demonstrate the need for leveraging spectral diversity to optimize overall system throughput and robustness,  whose study is enabled by the multi-frequency-band architecture of \arahaul.

\paragraph{\araoptical}
One major gap in the existing long-distance terrestrial FSOC studies is the lack of real-world measurement data on weather impact, that is, scintillation effect introduced by atmospheric turbulence. 

\begin{figure}[htb!]
  \centering
\includegraphics[width=0.7\textwidth]{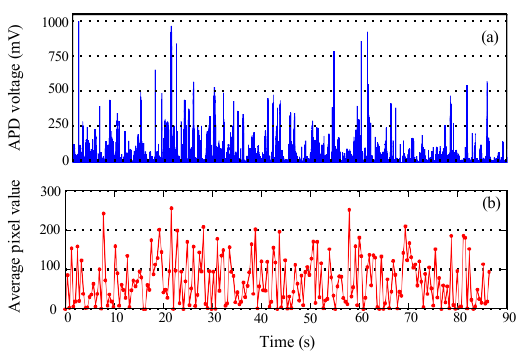}
  \caption{Example scintillation effect captured by the beacon receiver: (a) APD voltage over time; (b) average pixel value of the CMOS camera output over time. 
  }
  \label{fig:FSO_scin}
\end{figure}

\begin{figure}[h!]
    \centering
    \includegraphics[width=0.7\textwidth]{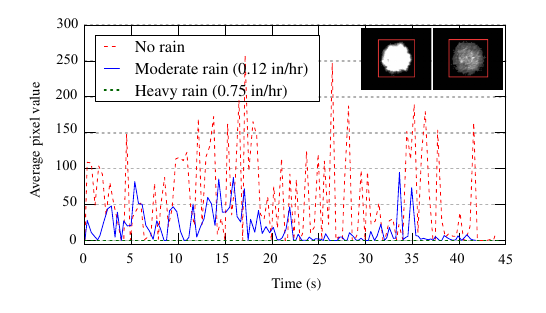}
    \caption{Comparison of the average pixel value (between 0 and 255) of CMOS camera output 
    under different rain conditions. The spot diagrams located at the upper right corner show the distorted beams due to scintillation: no rain~(left); moderate rain~(right). }~
    \label{fig:fsoc_rain_rate}
\end{figure}

Using the first-of-its-kind real-world \araoptical\ deployment and open 
APIs, we investigate the 
FSOC scintillation effect 
by measuring the receiver-side avalanch-photodiode (APD) voltage levels and the average pixel values of the receiver CMOS camera output. Selected results are shown in Figure~\ref{fig:FSO_scin}, 
where the temporal fluctuations of the APD voltage levels and pixel values over time can be clearly observed and precisely quantified. 
We also study how the scintillation effect varies under different rain conditions. As shown in Figure
~\ref{fig:fsoc_rain_rate}, 
the scintillation effect and losses become more severe under more intense rain conditions, and the impact can be precisely quantified too.
Such measurement studies will serve as a solid foundation for future research explorations, for instance, detailed characterization of FSOC channels under different weather conditions, 
and development of predictive models for FSOC control.

\subsubsection{Power Consumption Experiment}

Characterizing power consumption is critical for today's cellular networks for realizing the concept of green communication. With smart power distribution units~(PDUs), \ara\ provides monitoring and management APIs to characterize the power consumption of individual components in cellular networks. The APIs offer the following unique capabilities:
\begin{enumerate}
    \item Real-time power consumption monitoring for individual components in all BSes;
    \item Long-term power consumption data collection and storage in our database.
\end{enumerate}

\begin{figure}[htbp!]
  \centering
\includegraphics[width=0.7\textwidth]{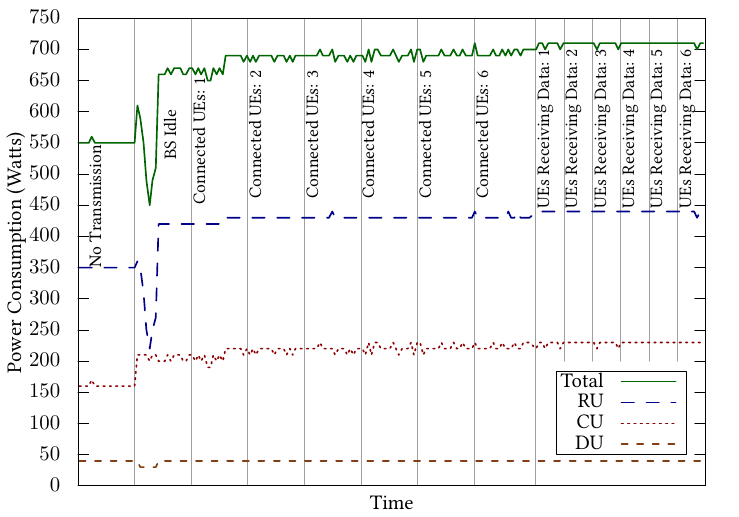}
  \caption{Power consumption of \aratvws\ BS in different states}
  \label{fig:pdu_skylark}
\end{figure}

Figure \ref{fig:pdu_skylark}
shows one such experiment where real-time power consumption pattern of the \aratvws\ BS under different states is measured. The states of the BS include \textit{no transmission, transmitting but idle, $k$ number of UEs connected,} and \textit{$k$ number of UEs transmitting data.} 
The power consumption increases up to 21\% from the first state (i.e., no transmission) to the last state (i.e., six UEs connected and sending data). There is a sudden decrease in power consumption when the BS start transmission which is attributed to the restarting of components in the individual RF chains in the frontend to make the BS ready for transmission. 

The long-term data collection capability allows analysis and modeling of power consumption over time. Table~\ref{tab:pdu_components} 
shows the power consumption (actual values and percentage) of individual components at \wilsonhall\ BS over a period of one month. We can observe that \aratvws\ has the most power consumption (since it is always operational) and \araoptical\ switches has the least. \arasdr\ has lower power usage because it is idle when not used in experiments. Such a comprehensive view of the distribution of power utilization across different equipment helps intelligently controlling the device ON/OFF times, thereby paving the path toward green communication.

\begin{table*}[!htbp]
\centering
\caption{Equipment-level Averaged Power Consumption (Watts/Amps) and Total Energy Consumed (kWh) Over 1 Month Period}
\label{tab:pdu_components}
\resizebox{\textwidth}{!}{
\begin{tabular}{cl|c|c|c|c|c|c|c|}
\cline{3-9}
\multicolumn{1}{l}{}                                       &       & \textbf{\aratvws} & \textbf{\arasdr} & \textbf{Compute} & \textbf{Switches} & \textbf{\araoptical} & \textbf{Other} & \textbf{Total} \\ \hline
\multicolumn{1}{|c|}{\multirow{2}{*}{\textbf{Value}}}      & Amps  & 5.822           & 4.377         & 5.370            & 2.332            & 1.392           & 0.783          & 20.08        \\ \cline{2-9} 
\multicolumn{1}{|c|}{}                                     & Watts & 692.2           & 415.2         & 647.2           & 188.1             & 115.5          & 45.2           & 2103.4       \\ \cline{2-9} 
\multicolumn{1}{|c|}{}                                     & kWh   & 498.4           & 298.9         & 466.0           & 135.4             & 83.2            & 32.5           & 1514.4       \\ \hline
\multicolumn{1}{|c|}{\multirow{2}{*}{\textbf{Percentage}}} & Amps  & 29.0\%          & 21.8\%        & 26.7\%          & 11.6\%            & 7.0\%          & 3.9\%          & 100\%          \\ \cline{2-9} 
\multicolumn{1}{|c|}{}                                     & Watts/kWh & 32.9\%          & 19.7\%        & 30.8\%          & 9.0\%             & 5.5\%           & 2.1\%         & 100\%          \\ \hline
\end{tabular}}
\end{table*}

\subsection{Example Rural Application Experiments 
}\label{subsec:appPilot}

\ara\ is architected to support wireless and application co-evolution such as
enabling higher-network-layer and  application research in addition to experiments at the lower network layers as shown in Section~\ref{subsec:wirelessExpt}. 
    In what follows, we demonstrate  experiments of an innovative wireless transport  solution integrated with a real-time, high-resolution video streaming application service as required by agriculture automation.

\paragraph{Real-Time Liquid Wireless Transport} 
Agriculture automation (e.g., real-time data analytics and teleoperation \cite{RT-LWN}) requires high-throughput, low-latency communications in the presence of uncertainties in wireless communications. To meet the demanding requirements, we implement a transport-layer solution called Liquid Transport Layer (LTL).
LTL uses the state-of-the-art RaptorQ fountain code \cite{liquid_video_stream, shokrollahi2011raptor} to improve network reliability while maintaining low latency by taking advantage of the \emph{liquid} nature of RaptorQ encoded data (specifically, with key properties of expandability and interchangeability~\cite{liquidDataNetworking}). The liquid nature enables low-overhead, reliable, and real-time data delivery by eliminating the delay in traditional, retransmission-based schemes \cite{RT-LWN}.

\paragraph{Experiment Setup} 
RaptorQ codes are \emph{rateless} and therefore able to generate as much redundant \emph{repair} data as necessary. LTL uses a value of 10\% for the repair data generation, therefore LTL is able to recover from packet loss up to a rate of 10\% (e.g., 10\,Mbps stream will be 11\,Mbps with LTL and be able to recover from up to 10\% packet loss). This value can increase or decrease via a feedback link, however we select 10\% as our default value. Furthermore, at the client, the LTL implementation utilizes upto 5\% CPU processing given a 40\,Mbps stream. 

We place a UE on an agriculture vehicle. The UE is connected to a $360\degree$ camera on the vehicle, which live-streams 4K 360\degree\ video at a bit rate of 30\,Mbps showcasing the potential of \ara\ to support high-throughput, real-time applications. 
The live video streaming 
is enabled by the \textit{GStreamer}~\cite{gst} multimedia framework, which is installed on the UE host computer. Using the GStreamer RTP plugin, the RTSP stream from the camera is pulled and sent to a streaming server located at the \ara\ data center.
The UE is equipped with an \arac\ radio, which is connected to the \wilsonhall\ BS. The vehicle moves along a designated route across a farm field, which is about 2.63\,km away from the BS. The speed of the vehicle is about 20\,km/hour and each experimental trial takes about 110\,seconds.
    We comparatively study the behavior of LTL and a baseline UDP solution without LTL.

\paragraph{QoE Metrics \& Measurement Results} 
We evaluate the Quality of Experience (QoE) in video streaming in terms of the following metrics: 1) \textit{Frames Per Second} (FPS): number of frames per second decoded at the receiver, 2)  \textit{Structural Similarity Index Measure (SSIM)}: comparison of structural information between the source 
video and received video, and 3) \textit{Stall Ratio}: percentage of time the receiver spends in waiting for new  data to update the live stream compared to the entire duration of the stream.

\begin{figure}[htb!]
  \centering
  \begin{subfigure}[b]{0.49\columnwidth}
    \centering
    \includegraphics[height=1.5in]{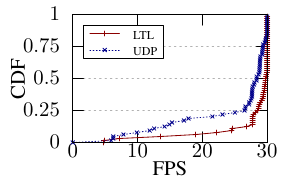}
    \caption{Good connectivity area}
    \label{fig:good_fps}
  \end{subfigure}
  \begin{subfigure}[b]{0.49\columnwidth}
    \centering
    \includegraphics[height=1.5in]{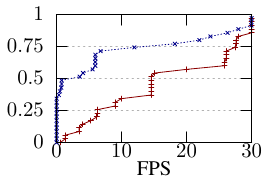}
    \caption{Bad connectivity area}
    \label{fig:bad_fps}
  \end{subfigure}
  \caption{Comparison of CDF of FPS for LTL and UDP video streaming solutions at the bit rate of 30\,Mbps}
  \label{fig:cdf_fps}
\end{figure}

Measurement results 
are shown in Figure~\ref{fig:cdf_fps} 
and Table~\ref{tab:ssim_stall}.
The \textit{good connectivity} area is defined as the region where the UE has a stable and high-throughput connection with the BS, while the \textit{bad connectivity} area refers to the edge of the cell with sporadic and unstable connectivity.
As shown in Figure~\ref{fig:good_fps}, LTL 
yields a high FPS close to 30 in the good connectivity area, and provides a higher overall FPS than the baseline UDP solution.
The improvement increases significantly as the UE moves away from the BS and gets closer to the edge of the cell, as shown in Figure~\ref{fig:bad_fps}.
Moreover,
LTL demonstrates a similar or higher mean SSIM score and a lower mean stall ratio than the baseline UDP solution, particularly in the bad connectivity area, as shown in Table~\ref{tab:ssim_stall}.

\begin{table}[h]
    \centering
    \footnotesize
    \caption{Comparison of SSIM Score and Stall Ratio (mean value $\pm$ 95\% CI) 
    for LTL and UDP solutions}
    \resizebox{0.8\textwidth}{!}{
    \begin{tabular}{c|c|c|c|c|}
    \cline{2-5}
    \multicolumn{1}{c|}{} & \multicolumn{2}{c|}{\textbf{Good Connectivity}} & \multicolumn{2}{c|}{\textbf{Bad Connectivity}} \\
    \cline{2-5}
    & \textbf{SSIM} & \textbf{Stall Ratio} & \textbf{SSIM} & \textbf{Stall Ratio} \\
    \cline{2-5} \hline
    \multicolumn{1}{|c|}{\textbf{LTL}} & 94.2\% $\pm$~0.3 & 3.8\% $\pm$~0.5 & 87.8\% $\pm$~0.3 & 41.8\% $\pm$~0.4\\
    \hline
    \multicolumn{1}{|c|}{\textbf{UDP}} & 95.0\% $\pm$~0.2 & 4.6\% $\pm$~0.3 & 76.5\% $\pm$~0.2 & 58.6\% $\pm$~0.2\\
    \hline
    \end{tabular}}
    \label{tab:ssim_stall}
\end{table}


\section{Related Work}
\label{sec:related-work}

While notable outdoor wireless testbeds exist worldwide, very few focus on rural and agriculture applications. In Japan, Niigata University has deployed a wireless mesh network testbed 
to address the digital divide in rural mountain areas~\cite{takahashi2007wireless}. 
    In Finland, the Converged Infrastructure for Emerging Regions (CIER) project has developed a wireless mesh network testbed for 
    robust, energy-efficient, and cost-effective heterogeneous wireless networking~\cite{CIER}. 
In the UK, the 5G RuralFirst~\cite{5GRuralF67online} testbed 
hosts a cloud core network and operates across an array of frequency bands including 700\,MHz, 2.4\,GHz, 3.5\,GHz, 5\,GHz, and 26\,GHz. It facilitates experimentation across diverse use cases encompassing dynamic spectrum sharing, broadcast, agriculture, and industrial IoT. 
    However, these testbeds do not support research in TVWS massive MIMO and long-distance, high-capacity x-haul, nor do they consider wireless and applicatoin co-evolution. 5G RuralFirst also does not have SDR systems, thus not supporting experiments in O-RAN and open-source NextG. 

In Europe, the 5G testbed at University of Bristol~\cite{harris2015BIO} is a part of the BIO (Bristol is Open) city testbed, with a specific focus on smart city applications. The NITOS~\cite{pechlivanidou2014nitos} testbed at 
University of Thessaly (Greece) has 50 nodes deployed on a building rooftop, 
supporting experiments with Wi-Fi, WiMAX, and LTE. However, they do not support experiments using SDRs, TVWS massive MIMO, and wireless x-haul systems, 
and NITOS does not support 5G experiments. 
    In the realm of wireless x-haul, the Patras 5G testbed~\cite{tranoris2019patras} hosts a private 5G network with a mmWave backhaul. However, it does not support experiments with 
    multi-modal, long-distance, and high-capacity wireless x-haul systems, nor does it support TVWS massive MIMO research. 
ExPECA \cite{expeca-james-kth} is an edge computing and wireless communication research testbed located inside a unique  isolated underground facility~(20m below ground) in Sweden, and is equipped with compute resources and SDR nodes, enabling wireless experiments inside a highly controlled environment.

Most closely related to \ara\ are the wireless testbeds from the U.S$.$ NSF Platforms for Advanced Wireless Research (PAWR) program \cite{PAWR}, that is, COSMOS~\cite{raychaudhuri2020cosmos}, POWDER~\cite{breen2020powder}, 
and AERPAW~\cite{marojevic2020aerpaw}.
    COSMOS and POWDER are deployed in the New York City and Salt Lake City, respectively. COSMOS focuses on mmWave communications, and POWDER focuses on programmable wireless networks. They do not focus on rural broadband, nor do they support wireless and applications research using COTS 5G massive MIMO systems. 
AERPAW is deployed in Research Triangle, North Carolina, and it focuses on UAV wireless networking. 
The COTS 5G platform of AERPAW does not support massive MIMO. None of these testbeds support research experiments in TVWS massive MIMO nor long-distance, high-capacity wireless and free-space optical x-haul. 


\section{Concluding Remarks}
\label{sec:concludingRemarks}

Addressing the unique challenges and opportunities of rural broadband, \ara\ is a first-of-its-kind wireless living lab featuring the state-of-the-art wireless access and x-haul platforms, as well as high-performance compute resources embedded end-to-end from user equipment to base stations, edge, and cloud. 
    \ara\ enables unique research experiments ranging from the modeling to the architectures, technologies, services, and applications of rural wireless systems, covering a wide range of topics such as those in O-RAN, open-source NextG, TVWS massive MIMO, NTN, spectrum innovation, integrated rural wireless access and x-haul, 360\degree\ video streaming, real-time edge data analytics, and agriculture automation. 
\ara\ enables wireless and application co-evolution through a testbed architecture that integrates fully-programmable SDR and configurable COTS platforms. One open challenge for future research is developing open-source NextG hardware and software platforms that meet stringent performance and robustness demands by real-world applications. 
    Leveraging \ara's support for research reproducibility and community collaboration, another area worthy pursuing is using \ara\ as a tool for building collaborative communities to address grand challenges in advanced wireless and rural broadband (e.g., predictable real-time wireless networking for safety-critical cyber-physical systems). 
Even though ARA has just completed its Phase 2 build-out, it has attracted more than 214 individual research users from 61 research teams at 38 organizations across academia, industries, non-profits, and governments in North America (i.e., U.S$.$ and Canada), Europe (e.g., Sweden, France), South America (e.g., Brazil), and Asia (e.g., Thailand). Readers interested in joining the ARA researcher community are welcome to check out \url{www.arawireless.org} for opportunities and contacts.

\section*{Acknowledgment}
This work is supported in part by the NSF awards 2130889, 2112606, 2212573, 2229654, and~2232461, NIFA award 2021-67021-33775, and PAWR Industry Consortium. 
    We thank John Bennett George for his contribution and support in ARA deployment, Miho Walczak for grant management, Alex Van Alstyne for equipment procurement and community engagement, all of the   partners and volunteers for their contributions to the ARA  design and implementation. 
    We also thank the colleagues at the PAWR Project Office 
    for their support and partnership.








\bibliographystyle{elsarticle-num}
\bibliography{references.bib}




\end{document}